\documentstyle[epsf,12pt]{article}

\newlength{\extralineskip}
\newcommand{\junk}[1]{}

\newcommand{\Z}{Z\!\!\!Z}
\newcommand{\R}{{\rm l\!R}}
\newcommand{\beq}{\begin{equation}}
\newcommand{\eeq}{\end{equation}}
\newcommand{\bea}{\begin{eqnarray}}
\newcommand{\eea}{\end{eqnarray}}
\newcommand{\bd}{\begin{displaymath}}
\newcommand{\ed}{\end{displaymath}}
\newcommand{\nn}{\nonumber}

\newcommand{\N}{{\cal N}}
\newcommand{\th}{{\rm th}}
\renewcommand{\S}{{\cal S}}
\renewcommand{\H}{{\cal H}}
\newcommand{\G}{{\cal G}}
\newcommand{\B}{{\cal B}}
\newcommand{\s}{\sigma}

\newcommand{\e}{{\rm e}}
\newcommand{\n}{{\widetilde n}}

\newcommand{\lb}{\{}
\newcommand{\rb}{\}}

\newcommand{\seq}{\setcounter{equation}{0}}
\begin{document}
%%%%%%%%%%%%%%%%%%%%%%%%%%%%%%%%%%%%%%%%%%%%%%%%%%%%%%%%%%%%%%%%%%%%%%%%%%%
\hfill NBI-HE-98-16
\vskip 15mm
\centerline{{\Large{\bf
Wilson Loops, Bianchi Constraints and Duality }}}
\centerline{{\Large{\bf in Abelian Lattice Models}}}
\vskip 2mm
\centerline{\bf Sebastian Jaimungal\footnote{email: jaimung@physics.ubc.ca. This work is supported in part by NSERC of Canada.}} \vskip 5mm
{\it
\centerline{The Niels Bohr Institute}
\centerline{Blegdamsvej 17}
\centerline{DK-2100 Copenhagen $\emptyset$, Denmark.}
\vskip 2mm
\centerline{and} \vskip 2mm
\centerline{Department of Physics and Astronomy}
\centerline{University of British Columbia} 
\centerline{6224 Agricultural Road} 
\centerline{Vancouver, British Columbia V6T 1Z1, Canada.}
}
\vskip 3mm

\begin{abstract}
\noindent
We introduce new modified Abelian lattice models, with inhomogeneous
local interactions, in which a sum over topological sectors are
included in the defining partition function. The dual models, on
lattices with arbitrary topology, are constructed and they are found
to contain sums over topological sectors, with modified groups, as in
the original model. The role of the sum over sectors is illuminated by
deriving the field-strength formulation of the models in an explicitly
gauge-invariant manner. The field-strengths are found to satisfy, in
addition to the usual local Bianchi constraints, global
constraints. We demonstrate that the sum over sectors removes these
global constraints and consequently softens the quantization condition
on the global charges in the system. Duality is also used to construct
mappings between the order and disorder variables in the theory and
its dual. A consequence of the duality transformation is that
correlators which wrap around non-trivial cycles of the lattice vanish
identically.  For particular dimensions this mapping allows an
explicit expression for arbitrary correlators to be obtained.
\end{abstract} 

\vskip 4mm


\newpage\setcounter{page}{1}

\section{Introduction}
\seq

In recent years various dualities have played an important role in
identifying the non-perturbative objects in string theories and a web
of connections between the perturbative string theories has been
conjectured. Strong-Weak duality ($S$-duality) has also shed light on
the strong coupling limit of $N=2$ supersymmetric gauge theories
\cite{SeWi94a,SeWi94b}.  However, the notion of duality has existed in
statistical models since the early work of Kramers and Wannier
\cite{KrWa41} on the two dimensional Ising model. In their pioneering
work duality was used to compute the critical temperature
at which the Ising model undergoes a second order phase
transition. $S$-duality has also been studied in the other spin models
\cite{Sa80} and in the context of gauge theories on the lattice
\cite{FrSu78}. Recently, duality in generalized statistical models
were studied on topologically non-trivial manifolds
\cite{GaJaSe98} and a method for constructing self-dual models was given. 
For other relevant work see \cite{Ra, DrWa82}.

The lattice models considered here have degrees of freedom which take
values in an Abelian group, $\G$, and live on the $(k-1)$-cells of the
lattice with local interactions. These are generalizations of well
known statistical models such as spin models and gauge theories. On
topologically non-trivial manifolds there are configurations of the
fields which have global charges around the non-trivial $k$-cycles of
the lattice. For example, on a torus spin models can have vortex
configurations which contains non-trivial flux around the two
canonical cycles of the lattice. Analogously, the statistical sum of a
gauge theory on $S^2\times S^2$ includes monopole configurations which
have non-trivial charge around the two two-sphere's. In the flat case
the vortices, or monopoles etc., must satisfy certain quantization
conditions (which depends on $\G$) around the elementary plaquettes , cubes, 
etc.~. On topologically non-trivial
manifolds the global charges must also be quantized. In
\cite{GaJaSe98} the global charges were forced to  satisfy either
the same quantization conditions as the local charges, or were
completely free of quantization. In the present work we will not make
such a restriction. This is implemented by introducing a sum over,
what we will term, topological sectors of the theory over a subgroup
of $\G$ (rather than the entire group, $\G$, or only its identity
element, as in \cite{GaJaSe98}).  As a note, summing over these
sectors is somewhat akin to introducing a sum over sectors in (super)
Yang-Mills to obtain the correct $2\pi$-periodicity which was
implemented in \cite{HaZh97} (see also \cite{ShVa88, KoSh97}).

We will demonstrate that even in these models the duality
transformations can be performed on arbitrary lattices. The
topological sectors are found to appear in the dual model, however,
the groups over which one sums over are modified through certain
duality relations which we will derive. To understand the role that
the sum over sectors play on the physical quantities in the model we
rewrite the model in terms of the more natural field-strength
variables \cite{Ha79,BaHa}. This rewriting shows that the sum over
topological sectors relaxes the quantization conditions on the global
charges which appear around the various $k$-cycles of the
lattice. This allows the introduction of fractional global charges
into the system. Using this new freedom we construct several self-dual
models which are qualitatively distinct from earlier examples in that
these models contain different fractional global charges around the
various $k$-cycles of the lattice. In addition, as a consequence of
duality we will prove that correlators which wrap around a non-trivial
$(k-1)$-cycle of the lattice vanish identically. It is also
demonstrated that if the dimension of the lattice and dimension of the
cells on which the interactions take place are equal, then correlators
are completely determined by the duality transformations. A particular
example is that one can compute an arbitrary Wilson loop on a
two-dimensional Riemann surface of a gauge theory, with the
topological modifications, directly from duality. Due to the existence
of a global fractional charge we find that the correlators are
invariant under what one denotes by the inside and outside of a Wilson
loop only if the representation of the loop respects the $n$-ality of
the fractional charge.  Similar results will hold for the
higher-dimensional analogs.

The outline of the paper is as follows. In section \ref{model} we
introduce some notations and the language of simplicial homology
required to define the models in their most natural form. The duality
transformations are then carried out on the general model. To close
the section several explicit examples of the models using standard
notation are given and their duals are constructed. In the next
section we rewrite the model in terms of the field-strengths,
dynamical variables which live on cells of one higher dimension than
the original degrees of freedom. This is performed in an explicitly
gauge invariant manner.  Such a rewriting leads to an understanding of
what the sum over topological sectors imply physically. We will
demonstrate that the sum relaxes the quantization conditions on the
global charges in the system and can be used to introduce fractional
charges around the various $k$-cycles of the lattice.  Section
\ref{fracmod} explores several examples of self-dual models which
contain these fractional charges. Section
\ref{CorrSec} is a detailed discussion of order and disorder
correlators, how duality relates them to one another, a proof that
certain correlators vanish identically due to topological constraints,
and an explicit computation of correlators in dimensions $d=k$.
Finally in section \ref{conclusion} we make our concluding
remarks. Throughout this work spin models and gauge theories are used
in writing down explicit examples, however, it should be noted that
it is straightforward to extend the examples to include higher
dimensional fields.

\section{The General Model and its Dual} \label{model}

It is possible to treat spin systems, gauge theories, anti-symmetric
tensor field theories and all higher dimensional analogs, under a
universal framework. This can be achieved if one restricts the
Boltzmann weights to have local interactions only. By local
interactions we mean the following: In a spin model the spins interact
with their nearest neighbours, and the Boltzmann weights are link
valued objects with arguments depending only the spins on the ends of
the link. For a gauge theory the dynamical fields live on the links of
the lattice, while the Boltzmann weight is a plaquette valued object
and depends on the product of the fields around the
plaquette. Anti-symmetric tensor field theories are restricted in a
similar manner, the Boltzmann weights are defined on the elementary
cubes of the lattice and depend on the product of the plaquettes
around that cube. etc.~In continuum language these models with local
interactions are field theories in which the dynamical variable is a
$(k-1)$-form, $\omega$, and the action depends solely on the exterior
derivative of that form, ${\rm d}\omega$. This definition is taken
even on manifolds with non-trivial topology. On the lattice, the
statement that the interactions are local amount to insisting that the
Boltzmann weights depends on the coboundary, $\delta \sigma$, of a
$(k-1)$-chain, $\sigma$.  This is the language of simplicial homology
and a precise definition of the coboundary operator will be given
shortly.  To be self-contained and to introduce our notations we give
a short review of this language in the next subsection.

\subsection{Mathematical Preliminaries} \label{math}

We now give a short review of simplicial homology (for more details
see e.g. \cite{Sp66}). Consider a lattice $\Omega$ and associate to
every $k$-dimensional cell of the lattice an oriented generator
$c_{k}^{(i)}$ where $i$ indexes the various cells of dimension $k$.
These objects generate the $k$-chain group, denoted by
$C_k(\Omega,\G)$, $$
\sum_{i=1}^{\N_k} ~g_i ~c_{k}^{(i)} = g \in C_k(\Omega,\G) \qquad,\qquad
g_i\in \G $$ Here $\G$ is an arbitrary Abelian group with group
multiplication implemented through addition and $\N_k$ is the number of
$k$-cells in the lattice $\Omega$.  An element $g\in C_k(\Omega,\G)$ is
called a $\G$-valued $k$-chain or simply a $k$-chain.  Clearly
$C_k(\Omega,\G) = \oplus_{i=1}^{\N_k} \G $.

Two homomorphisms, the boundary $\partial$ and the coboundary
$\delta$, define the chain complexes $(C_*(\Omega,\G), \partial)$ and
$(C_*(\Omega,\G), \delta)$ where $C_*(\Omega,\G)\equiv
\oplus_{k=0}^{d} C_k(\Omega,\G)$: \bea
0\!\stackrel{\partial_{d+1}}{\longrightarrow}\! C_{d}(\Omega,\G)\!
\stackrel{\partial_{d}}{\longrightarrow}\!\dots\!
\stackrel{\partial_{k+2}}{\longrightarrow} \!C_{k+1}(\Omega,\G)\!
\stackrel{\partial_{k+1}}{\longrightarrow}
\!C_k(\Omega,\G)\!\stackrel{\partial_k}{\longrightarrow} \!C_{k-1}(\Omega,\G)\!
\stackrel{\partial_{k-1}}{\longrightarrow} \!\dots
C_{0}(\Omega,\G)\!
\stackrel{\partial_{0}}{\longrightarrow} \!0 \nn\\ 0\!
\stackrel{\delta_{d}}{\longleftarrow} \!C_{d}(\Omega,\G)\!
\stackrel{\delta_{d-1}}{\longleftarrow} \!\dots\!
\stackrel{\delta_{k+1}}{\longleftarrow} \!C_{k+1}(\Omega,\G)\!
\stackrel{\delta_k}{\longleftarrow}
\!C_{k}(\Omega,\G)\!\stackrel{\delta_{k-1}}{\longleftarrow}
\!C_{k-1}(\Omega,\G)\!\stackrel{\delta_{k-2}}{\longleftarrow}
\dots C_{0}(\Omega,\G)\!
\stackrel{\delta_{-1}}{\longleftarrow} \!0 \nn \eea 
These homomorphisms are defined by their
actions on the generators $c_k^{(i)}$ (we display the dimension
subscripts on $\partial_k$ and $\delta_k$ only when essential), \beq
\partial c_k^{(i)} =\sum_{j=1}^{\N_{k-1}} [c_k^{(i)} : c_{k-1}^{(j)}]
~c_{k-1}^{(j)} ~~,~~ \delta c_k^{(i)} = \sum_{j=1}^{\N_{k+1}} [
c_{k+1}^{(j)}:c_{k}^{(i)}] ~c_{k+1}^{(j)}
\label{cobd} \eeq where the incidence number is given by,
$$ [c_k^{(i)} : c_{k-1}^{(j)}] = \left \{
\begin{array}{ll} 
\pm 1 & \mbox{if the $j^{\th}$ $(k-1)$-cell is contained in the $i^{\th}$
$k$-cell} \\ 0 & \mbox{otherwise}
\end{array}\right.$$ The plus or minus sign reflects the relative orientation of the
cells.  The boundary (coboundary) chains and the exact (coexact)
chains are defined as, 
$$
\begin{array}{lclclcl}
 B_k(\Omega, \G) & = & {\rm Im}~\partial_{k+1} & , 
&B^k(\Omega,\G) & = & {\rm Im}~\delta_{k-1} \cr
Z_k(\Omega, \G) & = &{\rm ker}~\partial_k & , 
&Z^k(\Omega,\G) &= &{\rm ker}~\delta_k 
\end{array}$$
These sets inherit their group structure from the chain complex.  The
boundary and coboundary operators are nilpotent: $\partial\partial =
0$ and $\delta \delta =0$.  The quotient groups, $$ H_k(\Omega, \G) =
Z_k(\Omega,\G)/B_k(\Omega,\G) \qquad , \qquad H^k(\Omega, \G) =
Z^k(\Omega,\G) / B^k(\Omega, \G) $$ are the homology and cohomology
groups respectively. The elements of the homology group are those
chains that have zero boundary and are themselves not the boundary of
a higher dimensional chain, while the elements of the cohomology group
are those chains that have zero coboundary and are themselves not the
coboundary of a lower dimensional chain.  In section \ref{IllExm} we
will give some explicit examples for the 2-tori.

We define the projection of a chain onto cell via the following
``inner product'',
$$ \langle
c_k^{(i)}, c_l^{(j)}\rangle =
\delta_{k,l}\;\delta^{i,j} $$ which is linear in both arguments.  The
boundary and coboundary operators are dual to each other with respect
to this operation,
\beq \langle \partial
c_k^{(i)}, c_{k-1}^{(j)} \rangle = \langle c_k^{(i)}, \delta
c_{k-1}^{(j)} \rangle \label{cbdual} 
\eeq 
If $\Omega$ is a triangulation of an orientable manifold,
then there exists the following set of isomorphisms between the homology
and cohomology groups,
\beq
H_k(\Omega, \Z) \cong H^{d-k}(\Omega, \Z) \cong H^{k}(\Omega, \Z) \cong
H_{d-k} (\Omega, \Z) \label{iso}
\eeq
We will denote the generators of the homology group by
$\{h_a:a=1,\dots,A_k\}$ and the cohomology group by
$\{h^a:a=1,\dots,A^k\}$. The isomorphisms, (\ref{iso}), imply the
following orthogonality relations for the generators,
\beq
\langle h_a , h^ b \rangle = \delta_a^b \label{cohodual}
\eeq
and $h_a$ is said to be dual to $h^a$, not to be confused with
interpreting on the dual lattice (to be defined shortly). The
isomorphisms also imply that $A_k=A^{d-k}=A^{k}=A_{d-k}$.

Let us work out a few illustrative examples of how this language is used. Let 
$\sigma \in C_0(\Omega, \G)$,  consider the following inner product,
$$
\langle \delta \sigma, c_1^{(l)} \rangle = \langle \sigma, \partial c_1^{(l)}
\rangle = \langle \sum_{i=1}^{\N_0} \sigma_i\, c_0^{(i)}, (c_0^{(l_1)} 
- c_0^{(l_2)}) \rangle = \sum_{i=1}^{\N_0} \sigma_i \,\langle
c_0^{(i)} , (c_0^{(l_1)} - c_0^{(l_2)}) \rangle = \sigma_{l_1} -
\sigma_{l_2} $$ where the link $l$ points from the site $l_1$ to the
site $l_2$. This is precisely the form of the interaction for a
nearest neighbour spin model. As another example take $\sigma \in
C_1(\Omega, \G)$ and repeat the above, $$
\langle \delta \sigma ,  c_2^{(p)} \rangle 
= \langle \sigma , \partial c_2^{(p)} \rangle
=\langle\sum_{l=1}^{\N_1} \sigma_l \,c_1^{(l)},\sum_{l'\in p} c_1^{(l')} \rangle
= \sum_{l=1}^{\N_1} \sigma_l \, \sum_{l'\in p} \langle c_1^{(l)}, c_1^{(l')}
\rangle 
= \sum_{l' \in p} \sigma _l $$ where $p$ labels the plaquette, and we
have assumed that the links are oriented consistently with that
plaquette so that no relative signs appear. Not surprisingly this reproduces
the form of the Wilson-like terms in a lattice gauge theory.

It will be necessary at some point during the analysis in the next few
sections to introduce the notion of the dual lattice. To obtain the
dual lattice one introduces a new chain group generated by a new set
of generators which are constructed from the generators of
$C_*(\Omega,\G)$. A generator of $C_*(\Omega^*,\G)$, $c_k^{*(i)}$, is
obtained from the generators of $C_*(\Omega,\G)$, $c_k^{(i)}$, by making the
following identifications:
\bea
c_k^{(i)} &\leftrightarrow& c_{d-k}^{*(l)} \\
~[ c_k^{(i)} : c_{k-1}^{(j)} ] &\leftrightarrow&
[ c_{d-k+1}^{*(j)} : c_{d-k}^{*(i)}]
\eea
In general this identification does not produce a lattice, however in
case the original lattice was a triangulation of a closed orientable
manifold it does and these are the cases we study here. Under this
identification one also finds that,
\bea
\langle \delta g , c_k^{(i)} \rangle &\leftrightarrow& \langle \partial g^*,
c_{d-k}^{*(i)} \rangle \nn\\
\langle \partial g , c_{k-2}^{(i)} \rangle &\leftrightarrow& \langle \delta g^*,c_{d-k+2}^{*(i)} \rangle
\eea
where $C_{k-1}(\Omega,\G)\in g=\sum_{i=1}^{\N_{k-1}}
g_i\;c_{k-1}^{(i)}$ and $C_{d-k+1}(\Omega^*,\G) \in g^* =
\sum_{i=1}^{\N_{k-1}} g_i \;c_{d-k+1}^{*(i)} =
\sum_{i-1}^{\N^*_{d-k+1}} g_i^* \; c_{d-k+1}^{*(i)}$. Consequently, 
the homology (cohomology) generators on the lattice $\Omega$ become
cohomology (homology) generators on the dual lattice $\Omega^*$.

\subsection{The Modified Abelian Lattice Theories}
 
Now that the terminology is fixed the models can be introduced in
their most general form.  Consider an orientable lattice which has
freely generated $k$-th cohomology group: $H^k(\Omega, \Z) \cong \Z
\oplus
\dots \oplus \Z $, and let $\lb h^a : a =1,\dots, A^{k}\rb$ denote its
generators. The model is defined through the following partition
function,
\beq
Z = \sum_{\{ n_a \in \H_a\} } \; \sum_{\s \in
C_{k-1}(\Omega, \G) } \; \prod_{p=1}^{\N_k} \; B_p \left( \left \langle
\left( \delta \s + n_a h^a \right) , c_k^{(p)} \right \rangle 
\right ) \label{MODEL}
\eeq
The groups $\H_a$ are in general subgroups of $\G$, and one must
define the addition in the Boltzmann weights to mean addition in the
group $\G$.  For example, if $\G=U(1)$ and $\H_a=\Z_N$ the addition
should be written as $\delta\s~+~(2\pi/N) n_a h^a$. In the case $\G$
and/or $\H$ are continuous the sums should be replaced by the
appropriate integrals. These models correspond to standard statistical
models when $\H_a$ are taken to be identity elements in $\G$. For
example, if $k=1$, (\ref{MODEL}) is a spin model with $\G$-valued
spins on the sites of a lattice interacting with their nearest
neighbour. For $k=2$, (\ref{MODEL}) is a gauge theory with gauge group
$\G$ and Boltzmann weight having arguments which depend on the sum
(since group multiplication is implemented additively here) of the
links that make up a plaquette. When $k=3$ the model consists of an
anti-symmetric tensor field where the arguments of the Boltzmann
weights depend on the sum over plaquettes in an elementary cube. The
role of the extra summations over $\{n_a\}$, which we will term
topological sectors of theory, will be explained in section \ref{BiId}
once the model is written in terms of field-strength variables. It
should be pointed out that some particular choices of the groups $\{
\H_a \}$ reduces to the cases studied in \cite{GaJaSe98} and
\cite{Ja98} and will be elaborated on in section \ref{IllExm}.

\subsection{The Dual Model} \label{DualModel}
 
In this section we will perform the duality transformations on
(\ref{MODEL}).  Although the method is fairly standard it is useful to
demonstrate how it is carried out in this language. The first step is
as always to introduce a character expansion for the Boltzmann
weights,
\beq 
B(g) = \frac{1}{|\G|} \sum_{r\in\G^*} b(r) \chi_{r}(g) ,\qquad 
b(r) = \sum_{g\in\G} \; B(g ) \; \chi_{r}(g)
\label{CharCoeff}
\eeq
where $|\G|$ is the order of the group and $\G^*$ denotes the group of
irreducible representations of $\G$, and inherits $\G$'s Abelian
structure (for example $\R^*=\R$, $U(1)^*=\Z$, $\Z_N^*=\Z_N$). In the
case $\G$, or $\G^*$, are continuous groups the normalized sum
over group elements should be replaced by the appropriate
integral. Throughout the remainder of this paper we will ignore
overall constants to avoid clutter in the equations. The character
expansion can be thought of as a Fourier transformation which respects
the group symmetries. It is interesting to note that the set of
irreducible representations of $\G$ form a group if and only if $\G$
is Abelian. This is the reason why it is difficult to derive duality
relations for non-Abelian statistical models and field theories.

Due to the Abelian nature of the group, and its representations,
the characters satisfy the following factorization properties,
\beq
\chi_r(g) \chi_s(g) = \chi_{r+s}(g) \quad,\quad \chi_r(g)\chi_r(g') = 
\chi_r(g+g')
\label{CharFactProp}
\eeq
In addition they satisfy the following orthogonality relations,
\bea
\sum_{g\in\G} \chi_r(g) {\overline \chi}_s(g) &=& \delta_{\G^*}(r-s) \nn\\
\sum_{r\in\G^*} \chi_r(g) {\overline \chi}_r(g') &=& \delta_{\G}(g-g') 
\label{CharOrtho}
\eea
where ${\overline \chi}_s(g)$ is the character of $g$ in the representation
conjugate to $s$, i.e. $\chi_{-s}(g)$, and the subscript on the
delta-function indicates that it is invariant under that group.

On inserting the character expansion (\ref{CharCoeff}) into the
partition function, (\ref{MODEL}) the sum over representations and
product over $k$-cells can be re-ordered.  This introduces a
representation, $r_p$, on every $k$-cell of the lattice, these
representations can be encoded into a $k$-chain denoted by $r\equiv
\sum_{p=1}^{\N_k} r_p c_k^{(p)}$.  The partition function
then becomes,
\beq
Z = \sum_{r\in C_k(\Omega, \G^*)} \prod_{p=1}^{\N_k} b_p(\langle r , c_k^{(p)}
\rangle )
\sum_{ \{ n_a \in \H_a\} } \; \sum_{\s \in
C_{k-1}(\Omega, \G) } \; \prod_{p=1}^{\N_k} \; \chi_{\langle r , c_k^{(p)}
\rangle }
\left( \left \langle
\left( \delta \s + n_a h^a \right) , c_k^{(p)} \right \rangle \right )
\label{Zdual1}
\eeq
Using the factorization properties of the characters, (\ref{CharFactProp}), 
and the definition of the (co)-boundary operator, (\ref{cobd}),
one can rewrite the product over characters in (\ref{Zdual1}),
\beq
\prod_{p=1}^{\N_k} \chi_{\langle r,c_k^{(p)}\rangle}\left(\left
\langle\left( \delta \s + n_a h^a \right) , c_k^{(p)} \right \rangle \right )
= 
\junk{\prod_{p=1}^{\N_k} \chi_{ \langle r , c_k^{(p)} \rangle}
\left( \left \langle \delta \s, c_k^{(p)} \right \rangle \right )
\prod_{p'=1}^{\N_k}
\chi_{ \langle r , c_k^{(p')} \rangle}
\left( \left \langle n_a h^a, c_k^{(p')} \right \rangle \right ) \nn\\
&&=
\prod_{p=1}^{\N_k} \chi_{ r_p}
\left(  \sum_{l=1}^{\N_{k-1}} [c_k^{(p)} : c_{k-1}^{(l)}] \;\s_l  
\right ) \prod_{p'=1}^{\N_k}\prod_{a=1}^{A^k} \chi_{r_{p'}}
\left(n_a \sum_{p''\in \gamma^a} \delta_{p',p''} \right)\nn\\
&&= \prod_{p=1}^{\N_k} \prod_{l=1}^{\N_{k-1}} \chi_{ r_p\;[c_k^{(p)} : 
c_{k-1}^{(l)}]} \left( \s_l \right) \prod_{p'=1}^{\N_k}\prod_{a=1}^{A^k}
\chi_{r_{p'}\sum_{p''\in \gamma^a} \delta_{p',p''} } \left(n_a\right) \nn\\
&&=
\prod_{l=1}^{\N_{k-1}} \chi_{\sum_{p=1}^{\N_k}r_p\;[c_k^{(p)} : c_{k-1}^{(l)}]}
\prod_{a=1}^{A^k} \chi_{\sum_{p'=1}^{\N_k}
r_{p'}\sum_{p''\in \gamma^a} \delta_{p',p''} } \left(n_a\right)
\nn\\
&&= }
\prod_{l=1}^{\N_{k-1}} \chi_{\langle \partial r , c_{k-1}^{(l)} \rangle}
\left( \langle \sigma , c_{k-1}^{(l)} \rangle \right)
\; \prod_{a=1}^{A^k} \chi_{\langle r , h^a \rangle} \left( n_a \right )
\label{bFlip}
\eeq
In the above the generators of the cohomology group were written as
$h^a =
\sum_{p\in\gamma^a} c_k^{(p)}$. The sum over $\sigma$ and $\{n_a\}$ in
(\ref{Zdual1}) can then be performed with use of the orthogonality
relations, (\ref{CharOrtho}),
\bea
\sum_{\{n_a \in \H_a \} } \sum_{\sigma} \dots &=&
\prod_{l=1}^{\N_{k-1}} \sum_{\sigma_l\in\G} \chi_{\langle \partial r , 
c_{k-1}^{(l)} \rangle} \left( \sigma_l \right) \;
\prod_{a=1}^{A^k} \sum_{n_a \in \H_a} \chi_{\langle r , h^a \rangle} 
\left( n_a \right ) \nn\\
&=& \prod_{l=1}^{\N_{k-1}} \delta_{\G^*} \left( \langle \partial r , 
c_{k-1}^{(l)} \rangle \right) \; \prod_{a=1}^{A^k} \delta_{\H_a^*} 
\left( \langle r , h^a \rangle\right) \label{rConstr}
\eea
The first constraint implies that the boundary of the $k$-chain $r$ vanishes
identically. The most general chain which satisfies that constraint is the
sum of a boundary of a higher dimensional chain and an element of the $k$-th
homology both carrying $\G^*$ coefficients:
\beq
r = \partial r' + h, \quad{\rm where}, \qquad r'\in C_{k+1}(\Omega, \G^*), \quad h \in H_k(\Omega,
\G^*) \label{rsol}
\eeq 
We will write the homology part of $r$ in terms of the generators of
the homology group, $h=\sum_{a=1}^{A_k} \n_a h_a$.  On inserting
(\ref{rsol}) into the remaining constraints in (\ref{rConstr}) and
using the orthogonality (\ref{cohodual}) one finds that coefficients
of the homology are further constrained,
\beq
\prod_{a=1}^{A_k} \delta_{\H_a^*} \left( \n^a \right) \label{hCoeff}
\eeq
 Since $\{\n^a\}$ are elements of $\G^*$ this constraint forces $\n^a
\in \G^*\perp\H^*_a$ (those elements in $G^*$ which are the identity in $H^*$)
where $\n_a$ is the coefficient of $h_a$ dual to $h^a$.  Inserting
(\ref{rsol}) and (\ref{hCoeff}) into the partition function
(\ref{Zdual1}) we obtain,
\beq
Z = \sum_{\{\n^a\in \G^*\perp\H_a^*\}} \;
\sum_{r\in C_{k+1}(\Omega, \G^*)} \; 
\prod_{p=1}^{\N_k} b_p\left(\left\langle \partial r + \n^a h_a , 
c_k^{(p)}\right\rangle \right) \label{Zdual2}
\eeq
where the dummy prime index on $r'$ from (\ref{rsol}) has been
removed. This representation of the model is not of the same form as
original one since the argument of the Boltzmann weight contains a
boundary operator acting on $r$ rather than a coboundary operator and
since $\{ \n^a \}$ are coefficients of homology rather than
cohomology. However, if one interprets the objects on the dual lattice
(see discussion at end of section \ref{math}) the partition function
reads,
\beq
Z = \sum_{\{\n_a\in \G^*\perp\H_a^*\}} \;
\sum_{r\in C_{d-k-1}(\Omega^*, \G^*)} \; 
\prod_{p=1}^{\N^*_{d-k}} b_p\left(\left\langle \delta r + \n_a h^{*a}, 
c_{d-k}^{*(p)}\right\rangle \right) \label{Zdual}
\eeq
Here $\{ h^{*a} :a = 1, \dots, A^{d-k}\}$ are the generators of
$H^{d-k}(\Omega^*, \Z)$ which are obtained by considering the
generators of $H_k(\Omega, \Z)$ on the dual lattice. Notice that this
final form of the dual theory has precisely the same form as
the original model (\ref{MODEL}). Clearly the following criteria are
necessary conditions for the model to be self-dual: the lattice must
be self-dual, $k$ must be equal to $d-k$ so that the dynamical degrees
of freedom are on the same types of cells and the coefficient group
$\G$ must be isomorphic to its group of representations
$\G^*$. Additional constraints on obtaining self-dual models are
imposed by the cohomology coefficients $\{\H_a\}$. The set of $\H_a$
must reproduce itself under duality, consequently every group
appearing in $\{\H_a\}$ must appear in $\{\G^*\perp\H_a^*\}$ with the same
multiplicity. In the next sub-section we will demonstrate how to
obtain some explicitly self-dual models using our results interpreted
in standard language.

\subsection{Illustrative Examples} \label{IllExm}

\begin{figure}
\epsfxsize=6in
\epsfbox[0 0 455 45]{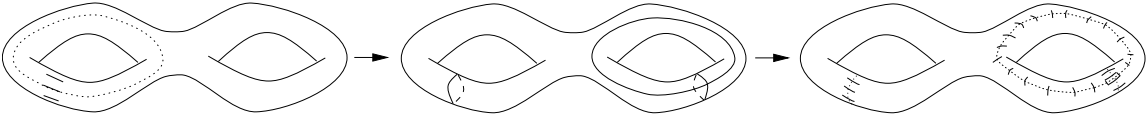}
\caption{Obtaining the dual theory to a spin model on surface of genus $g$ with
a sum over $\Z_2$ included over the cocycle in the first diagram. See
text for details.
\label{2tori}}
\end{figure}
As mentioned earlier these models are generalizations of the models
considered in \cite{GaJaSe98, Ja98}. Here we will make the connection
explicit and also introduce some examples which are unique to the
present discussion. Let us deal in particular with the case $k=1$, $\G
= \Z_2$ and $\Omega$ is a square lattice triangulation of a $2$-tori.
Since the dynamical variables live on the $(k-1)=0$-dimensional cells
of the lattice and interact through a nearest neighbour coupling this
corresponds to the Ising model with appropriate choice of Boltzmann
weights. The relevant cohomology here is $H^1(\Omega,\Z_2) \cong \Z_2
\oplus \Z_2 \oplus \Z_2\oplus \Z_2$. To understand what the generators of this group are it is best to consider the homology group of dimension $d-k=1$ on the
dual lattice. The generators of that group are well known: $h_a =
\sum_{l\in \gamma^*_a} c_1^{(l)}$ where $\gamma^*_a$ are the
set of links which wrap around the $a$-th handle of the 2-tori. The
cohomology generators of dimensions $k=1$ are then obtained by
interpreting each $\gamma^*_a$ back on the original lattice, denote
that set of links by $\gamma_a$. One visualization of $\gamma_a$ are
to view of them as ``steps of a ladder'' wrapping around the $a$-th
handle of the 2-tori. This construction is quite general: if the
lattice is orientable, first construct the generators of $H_{d-k}$ on
the dual lattice, and then interpret back on the original lattice that
will yield the generators of $H^k$.  We will consider the model in
which $\H_1 = \Z_2$ and all the rest are the identity element. The set
of links $\gamma_1$ are depicted in the first diagram of Figure
\ref{2tori} by the solid lines forming steps around the handle. The
model (\ref{MODEL}) in standard language is then given by,
\beq
Z = \sum_{n_1=\pm 1} \; \sum_{\{\sigma_i\}= \pm 1} \; 
\prod_{\langle i j \rangle\in\Omega} \; \exp\left\{ (n_1)^{\varepsilon(
{\gamma}_1;ij)} \; \beta \;\sigma_i \sigma_j \right\}
\eeq
here $i$ labels sites on the lattice, $\langle i j \rangle$ are
nearest neighbours forming a link and,
\beq
\varepsilon({\gamma}_1;ij) = \langle  \gamma_1, c_1^{(\langle i j\rangle)} 
\rangle = \langle \sum_{l\in\gamma_1} c_1^{(l)}, c_1^{(\langle i j\rangle)}
\rangle =
\left\{ 
\begin{array}{ll} 
1 & \mbox{ if $\langle i j \rangle$ is in ${\gamma}_1$} \\ 0 &
\mbox{otherwise}
\end{array} 
\right. 
\eeq
This model can be viewed as the sum of two Ising models: one which has
coupling constant $\beta$ on every link and another with coupling
$\beta$ on all links except for those contained in ${\gamma}_1$ where
it is taken to be $-\beta$.  Another view is to take the Ising
model on a $2$-tori and slice the lattice along one canonical cycle on
the dual lattice and then sew the lattice back together with the two
possible ``twists'' in the coupling.

Let us now demonstrate how the dual model is obtained. The first step
is to find the homology cycle which is dual, in the sense of
(\ref{iso}), to the generator $h^1=\sum_{l\in {\gamma}_1}
c_1^{(l)}$. The dual cycle, $h_1$, is depicted by the dotted line in
the fist diagram in Figure \ref{2tori}.  According to the analysis in
the previous sub-section the coefficient group of $h_1$, in the dual
theory before interpretation on the dual lattice (\ref{Zdual2}), is
$\G\perp\H = \Z_2\perp\Z_2=\{e\}$. Consequently, it is not summed over in the
dual theory. The remaining cycles, depicted in the second diagram in
Figure \ref{2tori}, have weight $\G\perp\{e\}= \Z_2\perp\{e\}=\Z_2$ and are
summed over in the dual theory although they were absent in the
original model. The final step is to interpret the groups associated
with the cycles $h_1,\dots,h_4$, as groups for the cocycles on the
dual lattice ( see the third diagram in Figure \ref{2tori}). Denote
these cocycles by $h_1^*, h_2^*, h_3^*$ (notice that $h_1^*$ is
isomorphic to $h_1$, it is simply shifted by $\frac 12$ lattice
spacing in both directions). The dual theory is then given by,
\beq
Z = c\left(\beta^*\right)
\left(\prod_{a=1}^3 \sum_{\n_a=\pm 1}\right) \; 
\sum_{\{\sigma_i^*\}= \pm 1} \; \prod_{\langle i, j \rangle \in \Omega^*} \; 
\exp\left\{ \left( \prod_{a=1}^3 (n_a)^{\varepsilon(\gamma^{*a};ij)} 
\right) \; \beta^* \;\sigma^*_i \sigma^*_j \right\}\label{notselfdual}
\eeq
where $\beta^*=-\frac 1 2 \ln \tanh \beta$ is the well known dual
coupling constant and the overall non-singular piece $c(\beta^*) = (2
\sinh (2\beta^*))^{-\frac {\N_1}{ 2}}$ appears when performing the
character expansion of the Boltzmann weight. Also, $\gamma^{*a}$
denotes the set of links which define $h^{*a}$. This particular
example is clearly not self-dual. However, it is not difficult to
convince oneself that if a sum over two cocycles, each in a separate
handle, was included in the original model, then the dual model would
be identical to the original one.  Thus on the $2$-tori there are $4$
inequivalent self-dual theories: $\{1,3\}$,$\{1,4\}$,$\{2,3\}$ and
$\{2,4\}$ where $\{i,j\}$ is intended to mean that a sum over the
cocycle labeled by $i$ and $j$ was included (we order the cocycles
so that 1 and 2 are around the first handle and 3 and 4 are around the
second). There are of course more self-dual models formed by summing
models of type (\ref{MODEL}). For example, a trivial way to obtain a
self-dual model is to take a model, which is not self-dual, and add it
to its dual (one could also subtract it from its dual to obtain
anti-self-dual models). This however, does not exhaust all possible
self-dual models that one can write down with $\H_a=\Z_2$ or the
identity. It is possible to find all of them by constructing a vector
of partition functions where the index labels all possible partition
functions, then carry out the duality transformation on each component
of the vector, to obtain a new vector of partition functions. Each
element of this ``dual'' vector must be a linear combination of the
original vector of partition functions since it contained all possible
partition functions. Consequently, one can construct a matrix which
acts on the original vector to give to the dual vector. This matrix is
a matrix representation of the duality transformation, and its
eigenvectors with eigenvalues plus one (it could have eigenvalues
minus one as well, since only its square must be the identity) label
the possible self-dual models. An example of this was given in
\cite{GaJaSe98} and we will not repeat this construction here.

The Ising model example only considered the degenerate case of $\H_a =
\G$ or the identity. It is, however, possible to relax this condition
and still obtain self-dual theories.  Consider the case of a spin
model on a $2$-dimensional Riemann surface and choose $\G=\R$ and
$\{\H_a = 2\pi R \Z\}$. Since the coefficient group is identical for
all cocycles, the dual theory is obtained by simply replacing $\G$ by
$\G^* = \G =\R$ and $\{\H_a \}$ by $\{\G^*\perp\H_a^* = R^{-1}
\Z\}$. This case was studied in \cite{Ja98} in the context of string
theory on a circle, the change in the coefficient $2\pi R \to R^{-1}$
is a manifestation of target space duality (T-duality). Other examples
with $\H_a=\H\ne\G$ were studied in \cite{Ja98} in the context of
string theory on discrete target spaces.

One further example of a self-dual spin-model on a Riemann surface is
furnished by the case $\G = \Z_{NM}$ and $\H_{i(a)} = \Z_N$ and
$\H_{i(b)} = \Z_M$ where $(a)$ indicates an $a$-cycle (the short
non-trivial cycle) and $(b)$ a $b$-cycle (the long non-trivial cycle)
on the surface, and $i$ labels the handles. We will delay a
description of this case until section
\ref{IllExm} where they will be interpreted as fractionally charged
models with different fractional charges around the various cycles of
the lattice.

\section{Local and Global Bianchi Identities} \label{BiId}
\seq

In pure gauge theories the action depends on the gauge fields only
through their field-strengths. It seems more natural then to rewrite
the theory so that the field-strengths are the dynamical degrees of
freedom. Various authors \cite{Ha79, BaHa, Ru97} have shown that the
price of doing this is the appearance of Bianchi constraints on the
field-strengths. On the lattice the gauge fields live on links, while
their field-strengths are defined on the plaquettes of the lattice.
The Bianchi constraints are realized, in this situation, by forcing
the group product of the field-strength variables in every elementary
cube to be the identity element.  In this section we consider the
generalized problem where the models are defined by the partition
function in (\ref{MODEL}) and we introduce a $k$-chain in place of the
coboundary of the $(k-1)$-chain $\s$. Consequently, the results of
\ref{constrsec} generalize the Abelian results found in \cite{Ha79, BaHa}. 
The variables which live on the $k$-dimensional cells will collectively
be referred to as field-strengths. We will show that the partition
function can be written in terms of field-strengths on topologically
non-trivial lattices. The usual Bianchi identities that arise in the
flat case are supplemented by constraints along the homology cycles of
the lattice as a direct consequence of the duality transformations
derived in the previous section. This will be demonstrated in an
explicitly gauge-invariant manner. The role of the sum over
topological sectors will also be illuminated by this rewriting of the
model.

\subsection{Computation of The Constraints} \label{constrsec}
 
It is possible to write down the natural choice of constraints by
counting degrees of freedom \cite{Ru97} which are introduced in the
field-strength formulation, however, we would like to derive them
explicitly. To do so, we make use of a Fadeev-Popov trick by inserting
the following identity\footnote{A similar trick was used for gauge
theories in the continuum \cite{Ha79} and gauge/spin models on the
lattice in \cite{BaHa}.} into the partition function (\ref{MODEL}),
\beq 1 = \sum_{f \in C_k(\Omega,
\G) } \; \delta_{\G} \left ( f - (\delta \sigma + n_a h^a) \right) 
\nn
\eeq
On insertion into (\ref{MODEL}) and re-ordering the summations one finds,
\beq
Z = \sum_{f \in C_k(\Omega, \G)} \;\prod_{p=1}^{\N_k} \; 
B_p \left( \left \langle f , c_k^{(p)} \right \rangle \right )
\Pi(f)
\label{FSform1}
\eeq
where $\Pi(f)$ is the constraint sum,
\beq
\Pi(f) \equiv \sum_{\{ n_a \in \H_a \} } \; 
\sum_{\s \in C_{k-1}(\Omega, \G) } \;
\prod_{p=1}^{\N_k} \delta_{\G}\left( \left\langle ( f - (\delta \sigma + n_a h^a), 
c_k^{(p)} \right \rangle \right)
\nn
\eeq
Notice that $\Pi(f)$ is itself a partition function of the form
(\ref{MODEL}) where the Boltzmann weights are $\G$-invariant delta
functions, the presence of the chain $f$ can be interpreted as an
external field acting on the system. The character coefficients of
such a Boltzmann weight are obviously,
\beq
b_p(r) = {\overline \chi}_{r} ( \langle f , c_k^{(p)} \rangle )
\nn
\eeq
The dual representation of $\Pi(v)$, before interpretation on the
dual lattice, is straightforward to write down and is given by
(\ref{Zdual2}),
\bea
\Pi(f) 
&=& \sum_{\{ \n^a\in \G^*\perp\H_a^*\}} \;
\sum_{r\in C_{k+1}(\Omega, \G^*)} \; \prod_{p=1}^{\N_k}
{\overline \chi}_{\langle \partial r + \n^a h_a, 
c_k^{(p)}\rangle}  \left( \langle f , c_k^{(p)} \rangle\right) \nn\\
&=& \left[ \sum_{r\in C_{k+1}(\Omega, \G^*)} \; \prod_{p=1}^{\N_k}
{\overline \chi}_{\langle \partial r, 
c_k^{(p)}\rangle}  \left( \langle f , c_k^{(p)} \rangle\right)\right] 
\!\!\left[ \sum_{\{\n^a\in \G^*\perp\H_a^*\}}
\prod_{p=1}^{\N_k} {\overline \chi}_{\langle \n^a h_a, 
c_k^{(p)}\rangle}  \left( \langle f , c_k^{(p)} \rangle\right)\right] \nn\\
&\equiv& \Pi_1(f) \; \Pi_2(f) \nn
\eea
The factorization properties (\ref{CharFactProp}) was used to obtain
the second equality.  This form of the constraints demonstrate that
they are two distinct classes of constraints: a topologically trivial
term $\Pi_1(f)$ and a term which is purely topological $\Pi_2(f)$.  We
will treat these two sectors separately. Using (\ref{bFlip}),
$\Pi_1(f)$ can be written in the following form,
\beq
\Pi_1(f) =\!\!\!\! \sum_{r\in C_{k+1}(\Omega, \G^*)} \prod_{c=1}^{\N_{k+1}}
{\overline \chi}_{\langle r, 
c_{k+1}^{(c)}\rangle}  \left( \langle \delta f , c_{k+1}^{(c)} \rangle\right)
= \prod_{c=1}^{\N_{k+1}} \!\!\delta_{\G} \left( \langle \delta f , 
c_{k+1}^{(c)} \rangle\right) 
= \prod_{c=1}^{\N_{k+1}}\!\! \delta_{\G} \left( \langle  f , 
\partial c_{k+1}^{(c)} \rangle\right)
\label{FreeCon}
\eeq
The orthogonality of the characters, (\ref{CharOrtho}), were used to
obtain the second equality. These constraints are the usual local
Bianchi constraints which forces the gauge connection to be flat
modulo harmonic pieces. Another interpretation of these constraints is
that within every $(k+1)$-cell a monopole like object can appear these
constraints force the charges to be quantized. For a spin model this
corresponds to the quantization of the vortex fluxes (the sum of the
vector field around a plaquette is quantized). For a gauge theory it
is the quantization of the monopoles charge (the sum of the
field-strength around an elementary cube is quantized). These latter
two examples were noted in the work of \cite{Ha79,BaHa}.

Let us now turn to the topological constraints. The factorization
properties, (\ref{CharFactProp}), and orthogonality relations,
(\ref{CharOrtho}), once again lead to a drastic simplification of the
expression,
\beq
\Pi_2(f) = \prod_{a=1}^{A_k} \sum_{n^a\in \G^*\perp\H_a^*} {\overline \chi}_{n^a} 
\left( \langle f , h_a \rangle\right) = \prod_{a=1}^{A_k} 
\delta_{(\G^*\perp\H_a^*)^*} \left( \langle f , h_a \rangle \right)
\label{TopCon}
\eeq
These constraints are the lattice analogs of holonomy constraints, in
the continuum they would correspond to constraints on the integral of
the field-strength around the canonical cycles of the manifold. We
will refer to these constraints as global Bianchi constraints as in
the original work of \cite{Ha79,BaHa} (which obtained similar results
for spin models and gauge theories without the use of topology and did
not include the sums over sectors). These constraints force
quantization conditions on the global charges around the
non-contractable $k$-cycles of the lattice.

As mentioned previously, the standard models correspond to choosing
$\{\H_a = \{e\} \}$, this implies that the local and global charges
satisfy the same quantization conditions. By making different choices
for $\{\H_a\}$ it is possible to introduce  fractional charges in
the system.  This will be discussed in detail in the next
section. Before closing this section, insert the two constraints,
(\ref{TopCon}) and (\ref{FreeCon}), into (\ref{FSform1}) to obtain the
full field-strength formulation of the model,
\beq
Z = \sum_{f \in C_k(\Omega, \G)} \prod_{p=1}^{\N_k} \; 
B_p \left( \left \langle f , c_k^{(p)} \right \rangle \right )
\;\prod_{c=1}^{\N_{k+1}} \delta_{\G} \left( \langle \delta f,c_{k+1}^{(c)} 
\rangle\right) \;\prod_{a=1}^{A_k} \delta_{(\G^*\perp\H_a^*)^*}
\left( \langle f , h_a \rangle \right) \label{FSform}
\eeq
Notice that if $d=k$ the local Bianchi constraints are absent and the
field-strength variables are constrained only through the global
constraints.  In section \ref{Corrdk} this fact will be used to obtain
explicit expressions for arbitrary correlators in dimension $d=k$.

\subsection{The Role of Summing Over Topological Sectors } \label{BIimp}

In section \ref{IllExm} we introduced several self-dual spin models by
making various choices of the groups $\G$ and $\{\H_a\}$.  In
particular consider the example where $\Omega$ is a square
triangulation of an orientable two-manifold of genus $g$, $\G=\Z_2$
and $\H_a=\Z_2$ for the $a$ cocycles (the generators which wrap
around the handles ``vertically'') and $\H_a=\{e\}$ for the $b$
cocycles (the generators which wrap around the handles
``horizontally''), also choose the standard Ising Boltzmann weights.
To obtain the field-strength formulation we must find the cycles dual
to the cocycles which are summed over and impose the
$(\G^*\perp\H_a^*)^*$ constraints on them. Firstly, consider the $b$
cocycles, the $a$ cycles are dual to these and thus have
$(\G^*\perp\H_a^*)^*=(\Z_2\perp\{ e\})^*=\Z_2$ constraints on the coefficient
group. The $a$ cocycles are dual to the $b$ cycles, consequently the
constraints on those cycles will be $\Z_2\perp\Z_2 =\{e\}$, i.e. there are
no constraints along those cycles.  The partition function can then be
written in terms of the field-strengths as follows,
\beq
Z =\sum_{\{f_l = \pm 1
\}} \prod_{\langle i  j\rangle\in\Omega}
\e^{ \beta f_{\langle i j\rangle}} \; 
\prod_{\Box\in\Omega} \frac {1}{2}\left(1 + \prod_{l\in\Box} f_l \right) \;
\prod_{a=1}^g \frac{1}{2} \left( 1 + \prod_{l\in\gamma_{2a}} f_l \right)
\label{vortex}
\eeq
The explicit form of a $\Z_2$-invariant delta function was used here.
Recall that $\gamma_a$ are the set of $k$-cells which define the
generator of the homology group. In this instance $\gamma_a$
correspond to the set of links forming the $a$-th non-contractable
loop around the handles of the lattice ($a=1,\dots,2g$).  The set
$\{\gamma_{2a}: a=1,\dots,g\}$ contain the $a$ cycles, and
$\{\gamma_{2a-1}: a=1,\dots,g\}$ contain the $b$ cycles.  The
constraints in (\ref{vortex}) imply certain restrictions on the
vortices in the model. The local Bianchi constraints force the vortex
flux within each elementary plaquette to vanish and the global Bianchi
constraints force the global vortex flux to vanish. However, since the
global Bianchi constraints are only present for the $a$ cycles,
arbitrary vortex configurations around the $b$ cycles are allowed
while the vortex flux must vanish around the $a$ cycles. This example
serves to illustrate that including a sum over the entire group along
a particular cocycle removes the constraints on the global charges
which wrap around its dual cycle.

As a second illustration, consider the same lattice, but with groups
$\G=\R$ and $\H_a =\Z$ for all cocycles. This model was also
demonstrated to be self-dual in section \ref{IllExm}. In terms of
field-strengths it is clear that since all cocycles have the same
group then the global Bianchi constraints force the global vortex
fluxes to respect  $(\R^*\perp\Z^*)^* = U(1)$ constraints. The partition
function reads,
\beq 
Z=\left(\prod_{\langle i j\rangle} \int {\rm d} f_{\langle i  j\rangle }\right)
\prod_{\langle i  j\rangle \in \Omega} B_{\langle i  j\rangle }
\left( f_{\langle i  j \rangle} \right) \;
\prod_{\Box\in\Omega} \delta \left( \sum_{l\in\Box} f_l \right) \;
\prod_{a=1}^{2g} \sum_{m_a\in\Z}\delta\left( 2\pi m_a - \sum_{l\in\gamma_a} f_l
\right) \label{circle}
\eeq
The local Bianchi constraints imply that no vortex fluxes are allowed
in the elementary plaquettes, while the global Bianchi constraints
forces the vortex fluxes around the cycles of the lattice to be
quantized in units of $2\pi\Z$. This illustrates that when the
topological sectors are summed over a subgroup of the group in which
the dynamical variables take values in, then the global charges that
arise are not completely free, but rather they are forced to satisfy a
relaxed quantization condition.

As a final example consider the case of a $U(1)$ gauge theory on the
lattice $\Omega$. In the present language this model is obtained by
choosing $\G=U(1)$ and $\{\H_a = \{e\} \}$. The field-strengths in
this case are dynamical plaquette valued fields. The local Bianchi
constraints force the sum of the field-strength around each
elementary cube to be $2\pi \times$ integer, and correspond to the
quantization of the monopole charges.  Since $(U(1)^*\perp\{e\})^*= U(1)$,
the global Bianchi constraints force the sum of the field-strength
around all two-cycles of the lattice to be $2\pi\times$ integer as
well, and corresponds to the quantization of the global
charges. Consequently in a standard gauge theory all topological
excitations are quantized by the same fundamental charge, as one would
expect. Now consider the case where one chooses $\H_1=U(1)$, and
the remaining $\H_a = \{e\}$. Then the local charges remain quantized
in units of $2\pi\Z$, while the global charge around the canonical
two-cycle dual to the cocycle $h^1$ is completely free of constraints
since $(U(1)^*\perp U(1)^*)^*=\{e\}$. Clearly, if more than one sector is
summed over the result simply generalizes. Let $\{\H_a= U(1) : a \in B
\subset 1,\dots,A^k\}$ and $\H_a = \{e\}$ otherwise, then using
(\ref{FSform}) the model in field-strength form is,
\beq
Z = \prod_{p=1}^{\N_2} \int_{-\pi}^\pi \frac{{\rm d}f_p}{2\pi} \; 
B_p \left(  f_p \right )
\;\prod_{c=1}^{\N_{3}} \sum_{m\in\Z} 
\delta \left( 2\pi m - \sum_{p\in c} f_p \right)
\prod_{a\notin B} \sum_{m'\in\Z} \delta \left(2\pi m'-\sum_{p\in\gamma_a}
f_p \right) \label{liberate}
\eeq
In this case $\gamma_a$ is the set of plaquettes which wrap around the
$a$-th two-dimensional hole in the lattice.  The above form of the
model should make it clear that including a sum over a particular
cocycle removes the quantization condition on the charges which are
enclosed by the cycle dual to it much like the liberation of the
vortex quantization earlier.

One question remains: what happens if a sum over a proper subgroup of
the full group, i.e. $\H_a \lhd \G$, was included?  Continuing with
the example of $U(1)$ gauge theory, take $\{ \H_a =
\Z_{N_a} \}$. Then the topological charge around the canonical
two-cycle dual to the cocycle $h^a$ can contain a fractional charge
$2\pi/N_a~\times$ integer while the local charges are still quantized
in units of $2\pi \Z$. To see this notice that $(U(1)^*\perp\Z_{N_a}^*)^*
= (\Z\perp\Z_{N_a})^* = (N_a)^{-1} \Z$, so that the field-strength
formulation of this model is,
\beq
Z = \prod_{p=1}^{\N_2} \int_{-\pi}^\pi \frac{{\rm d}f_p}{2\pi} \; 
B_p \left(  f_p \right )
\;\prod_{c=1}^{\N_{3}} \sum_{m_c\in\Z} 
\delta \left( 2\pi m_c - \sum_{p\in c} f_p \right)
\prod_{a=1}^{A_2} 
\sum_{m'_a\in\Z} \delta \left(\frac{2\pi}{N_a} m'_a-\sum_{p\in\gamma_a}
f_p \right)
\eeq
We will use the fact that one can introduce different fractional
charges around the various cycles of the lattice to construct new
self-dual models in the next section.

\section{Fractionally Charged Self-Dual Models} \label{fracmod}
\seq

In this section we will construct self-dual models which contain
several distinct fractional global charges. It will assume that the
Boltzmann weight is chosen so that its character coefficients have
the same functional form as the Boltzmann weight itself. This allows
us to focus on the set of $\{\H_a\}$ which lead to self-dual models.
It was shown in section \ref{IllExm} that on a genus $g$ surface, if
$\H_a=\G$ for all $a$ cycles then a spin model on that surface is
explicitly self-dual. In section \ref{BIimp} we pointed out that
introducing such a sum over the entire group releases the quantization
condition on the topological charge around that cycle (see
Eq.(\ref{vortex})). It was also demonstrated in section \ref{IllExm}
that the model with $\G=\R$ and $\H_a = \Z$ for every cycle on the
surface was self-dual. In this case the local charges are forced to
vanish, while the global charges were quantized in units of $2\pi\Z$
(see Eq.(\ref{circle})).  The natural question to ask is whether one
can construct a self-dual model which has different global charges
around the various generators. In the next subsection we construct a
collection of spin models in two-dimensions which have such a
distribution of fractional charges. In addition, we will construct
gauge theories in four dimensions which have different fractional
charges around the various two-cycles of the lattice.

\subsection{Spin Models}

\begin{figure}
\epsfxsize=6in
\epsfbox[0 0 461 112]{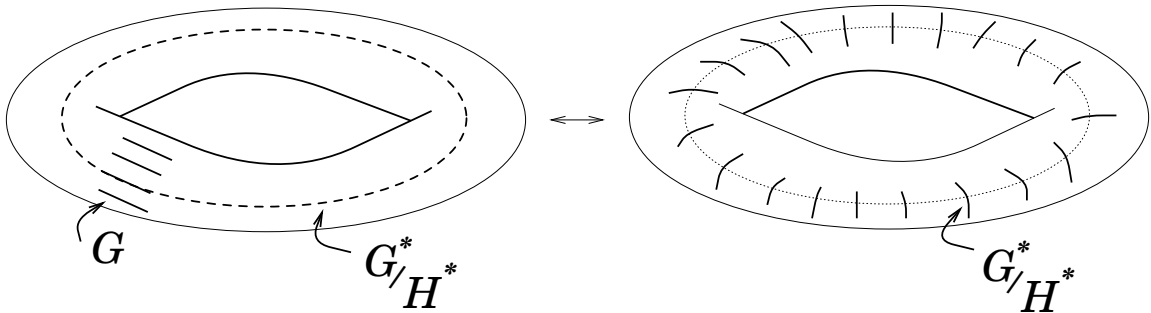}
\caption{Constraints on the cocycles in the first diagram force constraints
on the dual cycle (shown by the dotted line), interpreting on the dual
lattice leads to constraints on the cocycle in the second diagram.
\label{dual1tori}}
\end{figure}

\begin{table}
\begin{center}
\begin{tabular}{|r|c|c|c|} \hline
& Spin Variable & Coefficient of $h^1$ & Coefficient of $h^2$ \\ \hline
Original Model & $\G$ & $\H_1$ & $\H_2$ \\ \hline
Dual Model  & $\G^*$ & $\G^*\perp\H_2^*$ & $\G^*\perp\H_1^*$ \\ \hline 
\end{tabular}
\caption{Mixing of the coefficient group under duality for a spin model on 
a torus.\label{mix12}}
\end{center}
\end{table}
As a first example consider a spin model on a torus with $\G=\Z_P$,
this forces $\H_1 = \Z_M$ and $\H_2=\Z_N$ for some integers $N$ and
$M$.  It is clear that in order for $\H_a$ to act freely on $\G$, $N$
and $M$ must be factors of $P$. Table \ref{mix12} contains the
transformations of the coefficient groups under duality. It is obtained
as follows, let $\{h^1, h^2\}$ be the generators of $H^1(\Omega,\G)$
(see Figure \ref{dual1tori}), $h^1$ consists of the set of links
wrapping around the short direction in the first diagram, and $h^2$ is
the set of links wrapping around the long direction in the second
diagram. According to the dual construction: the dual to $h^1$, $h_1$,
is forced to have coefficient group $\G^*\perp\H_1^* =
\Z_{P/M}$. Interpreting $h_1$ on the dual lattice leads to an object
which is isomorphic to the generator $h^2$ (see Figure
\ref{dual1tori}). Repeat the above, starting with the generator $h^2$
and combining the two results imply that under duality $\H_1= \Z_M \to
\Z_{P/N}$ and $\H_2 = \Z_N \to \Z_{P/M}$. Searching for self-dual
models then leads to two linearly dependent equations, which leaves two
of the three variables $M,N,P$ undetermined. It is clear that $P=MN$
is the most general solution which leaves $\H_a$ invariant. The partition
function in terms of field-strengths is easily written down,
\bea
Z &=& \sum_{\{ f_l = 0,\dots,MN-1\}} 
\prod_{\langle i , j \rangle\in\Omega}B\left( f_{\langle i,j\rangle} \right) \;
\prod_{p=\Box\in\Omega}\;\sum_{m_p \in \Z}
\delta\left( 2\pi m_p MN - \sum_{l\in\Box} f_l\right) \nn\\
&& \hspace{5mm}
\left( \sum_{m'_1 \in \Z} \delta\left( 2\pi m_1' M
- \sum_{l\in\gamma_1} f_l \right) \right)
\left( \sum_{m'_2 \in \Z} \delta\left( 2\pi m_2' N
- \sum_{l\in\gamma_2} f_l \right) \right)
\eea
Consequently, the model has local charges which are quantized in units
of $2\pi MN$ while the global charge around the cycle $h_1$ is
quantized in units of $2\pi M$ and around the cycle $h_2$ is quantized
in units of $2\pi N$. This is the first illustration of a model in
which the global charges are unrelated to one another, yet the model
is explicitly self-dual. This idea can be generalized to the case of
an orientable surface of genus $g$. Simply choose $\G=\Z_{NM}$ and
$\H_a = \Z_{N}$ for the $a$ cycles and $\H_a = \Z_M$ for the $b$
cycles.  This choice is invariant since starting with an $a$ cocycle
($b$ cocycle) and then imposing the constraints on the dual $b$ cycle
($a$ cycle) and finally interpreting the cycle on the dual lattice
leads to a cocycle isomorphic to the $b$ cocycle ($a$ cocycle)
around the same handle as the original $a$ cocycle ($b$ cocycle). In
other-words the duality transformations only serve to mix the groups
within each handle. Since we have proven that the case of a single
handle can be made duality invariant, this argument demonstrates that
the model on the genus $g$ surface is also duality invariant.

\subsection{Gauge Theories}

For gauge theories a similar analysis can be carried out. In this case
the dimension of the lattice has to be four. We will consider two
lattices which produce self-dual models with multiple fractional
charges, there are of course many more. The first example is a lattice
with the topology of $S^2 \times S^2$. Firstly $H_2(\Omega,\G)\cong \G
\oplus \G$ and the generators are given by the sets of plaquettes
which encases one of the two-spheres. The generators of
$H^2(\Omega,\G)$ are then the plaquettes which are dual to those
surfaces, which can be thought of as plaquettes perpendicular to the
two-sphere (of course this is in four dimensions). Table \ref{mix12}
still holds in this situation, where $h^1$ and $h^2$ are taken to be
the appropriate generators, and $\G$ is now the gauge-group rather
than the spin group. Consequently, as in the last section, choosing
$\G=\Z_{NM}$, $\H_1 = \Z_M$ and $\H_2=\Z_N$ leads to an explicitly
self-dual model containing distinct fractional charges around the two
cycles of the lattice.

\begin{table}
\begin{center}
\begin{tabular}{|r|c|c|c|c|c|c|c|} \hline
&  & \multicolumn{6}{c|}{Coefficient Group of} \\ \hline
& Gauge Group & $h^1$ & $h^2$ & $h^3$ & $h^4$ & $h^5$ & $h^6$ \\ \hline
Original Model & $\G$ & $\H_1$ & $\H_2$ & $\H_3$ & $\H_4$ & $\H_5$ & $\H_6$\\ 
\hline
Dual Model  & $\G^*$ & $\G^*\perp\H_2^*$ & $\G^*\perp\H_1^*$ & $\G^*\perp\H_4^*$ & 
$\G^*\perp\H_3^*$ & $\G^*\perp\H_6^*$ & $\G^*\perp\H_5^*$\\ \hline 
\end{tabular}
\caption{Mixing of the coefficient group under duality for a gauge theory
on $T^4$. See the text for description of $h^i.$\label{mix6}}
\end{center}
\end{table}
As a second example consider the lattice with the topology of
$T^4=S^1\times S^1\times S^1 \times S^1$. In this case there are six
generators of the homology and cohomology: $H^2(\Omega,\G)\cong
H_2(\Omega,\G) \cong \oplus_{a=1}^6 \G$. The homology generators are
the plaquettes which make up the $4\choose 2$ possible tori
constructed out of the four circles, and as usual the cohomology are
the dual to those. Six coefficient groups must then be specified to
define the model. However, just as in the spin model on the genus $g$
surface, the generators form pairs, which under duality only
communicate within each pair. This feature is a result of
Poincare-duality.  For example, the cocycle corresponding to the torus
formed by the first two circles will force the coefficient group of
the cycle corresponding to the torus formed by the last two circles to
be $\G^*\perp\H_1^*$. Interpreting this cycle on the dual lattice leads to
a cocycle which is isomorphic to the generator corresponding to the
torus formed by the last two circles. Consequently, the transformations
of the coefficient group can be summarized as in Table \ref{mix6}
where the cohomology generators $h^i$ are dual to the following
generators of the homology: $h_1 \sim \{1,2\}$, $h_2 \sim \{3,4\}$,
$h_3 \sim
\{1,3\}$, $h_4\sim\{2,4\}$, $h_5\sim\{1,4\}$ and $h_6\sim\{2,3\}$. The
notation $h_a\sim\{i,j\}$ means that two-cycle consists of plaquettes
which wrap around the torus formed by the $i$-th and $j$-th circle. It
should be clear from the examples that under the duality $h^{2a}$
($h^{2a-1}$) forces constraints on $h_{2a+1}$ ($h_{2a}$) and then
interpretation on the dual lattice implies that constraints are forced
on $h^{2a-1}$ ($h^{2a}$) and one finds the situation depicted in Table
\ref{mix6}.  At the level of the coefficient groups, the present case
is identical to the spin model on a surface of genus 3. This implies
that choosing $\G=\Z_{MN}$, $\H_{2a-1} = \Z_N$ and $\H_{2a} = \Z_M$,
$a=1,2,3$ (one could also mix $N$ and $M$ for different values of $a$)
leads to self-dual models with different fractional charges around the
various cycles of the lattice.

These examples do not by any means exhaust the possible models which
exhibit self-duality with several fractional charges. They do serve to
illustrate the canonical situations in which they appear however. It
is not entirely clear what physical situations in which such models
would arise, however, the fact that they are explicitly self-dual
warrants some interest on its own.

\section{Correlators} \label{CorrSec}
\seq

Duality not only relates two partition functions to one another, it
also relates the disorder and order correlators in the theory and its
dual. The focus of this section will be to obtain expressions for the
correlators in the original model in terms of correlators in the dual
theory. As a consequence of this rewriting, correlators in models in
dimension $d=k$ can be computed explicitly. Also, a vanishing
correlator theorem will be proven for a certain subset of correlators.

An order or disorder correlator is defined through the following statistical
sum,
\bea
\left\langle \prod_{p\in\gamma} \chi_s \left(\langle \sigma,\partial c_k^{(p)} 
\rangle \right) \right \rangle
&\equiv&
\frac 1 Z \sum_{\{ n_a \in \H_a \}}
\sum_{\s \in C_{k-1}(\Omega, \G) }\;
\prod_{p\in\gamma}\chi_s\left(\langle\sigma, \partial c_{k}^{(i)}\rangle\right)
\nn\\
&&\hskip 38mm \prod_{p=1}^{\N_k} \; B_p \left( \left 
\langle \left( \delta \s + n_a h^a \right) , c_k^{(p)} \right \rangle \right )
\label{correlator}
\eea
where $Z$ is the partition function (\ref{MODEL}), and $\gamma$ is a
collection of $k$-cells on the lattice. Particular choices of $\gamma$
lead to the order or disorder correlators.  

The order correlator is defined by choosing $\gamma$, on the original
lattice, such that it spans an arbitrary arc-wise connected
$k$-dimensional surface with boundary. Define the chain $\Gamma \equiv
\sum_{p\in\gamma} c_k^{(p)}$, then consider the boundary of this chain
$\partial \Gamma =
\sum_{l\in\partial\gamma} c_{k-1}^{(l)}$. In general this will consist
of several disconnected components, denote this set of boundaries by
$\{ \B_i = \sum_{l\in(\partial\gamma)_i} c_{k-1}^{(l)}\}$ with
$i=1,\dots,b$.  Then (\ref{correlator}) is the order correlator of
$\{\B_i\}$. One can see this explicitly by using the factorization
properties of the characters, (\ref{CharFactProp}), to show that,
\beq
\prod_{p\in\gamma} \chi_s \left(\langle \sigma,\partial c_k^{(p)} 
\rangle \right) 
= \chi_s \left( \langle \sigma, \partial \Gamma \rangle \right) =
\prod_{i=1}^b \chi_s \left( \langle \sigma, \B_i \rangle \right) =
\prod_{i=1}^b \chi_s \left( \sum_{l\in(\partial\gamma)_i} \sigma_l \right)
\label{ordersimp}
\eeq
Consider the case of a spin model ($k=1$), one
example of an order correlator is given by taking $\gamma$ to be a set
of links running between site $i$ and site $j$ then the correlator is,
$$
\left\langle
\prod_{p\in\gamma} \chi_s \left(\langle \sigma,\partial c_k^{(p)} 
\rangle \right) \right \rangle
= \left\langle \chi_s( \sigma_i ) \chi_{-s} (\sigma_j) \right\rangle
$$ which is the usual two-point function between sites $i$ and $j$.

Disorder correlators are defined by considering a correlator in the
dual model and then re-interpreting it back on the original
lattice. Let $\gamma^*$ be a collection of $(d-k)$-cells on the dual
lattice (notice that $d-k\ge 0$ since the models we consider interact
on the $k$-cells of the lattice) which forms an arc-wise connected
$(d-k)$-dimensional surface with boundary. This could be used to
define a correlator in the dual theory.  Instead consider the
collection of $k$-cells on the original lattice which are dual to
these $(d-k)$-cells, denote this collection by $\gamma$. It should be
clear that $\gamma$ does not form an arc-wise connected
$k$-dimensional surface, rather it forms a collection of disconnected
$k$-cells.  Let $\{\B^*_i\}$ be the collection of $(d-k-1)$-cells
which form the boundaries of $\gamma^*$ on the dual lattice, and
denote by $\B_i$ the collection of $(k+1)$-cells on the original
lattice dual to $\B^*_i$. The correlation function (\ref{correlator})
with the above choice for $\gamma$ is the disorder correlator of the
collection of $(k+1)$-cells $\{B_i\}$. It is possible to obtain
$\{\B_i\}$ directly from $\gamma$ since $\{\B^*_i\}$ are the
boundaries of $\gamma^*$ on the dual lattice and $\{\B_i \}$ are the
set of $(k+1)$-cells forming the coboundary of $\gamma$ on the
original lattice. There is no simplification of the disorder
correlators as there was for the order correlators (\ref{ordersimp}).

\begin{figure}
\epsfxsize=6in
\epsfbox[0 0 443 330]{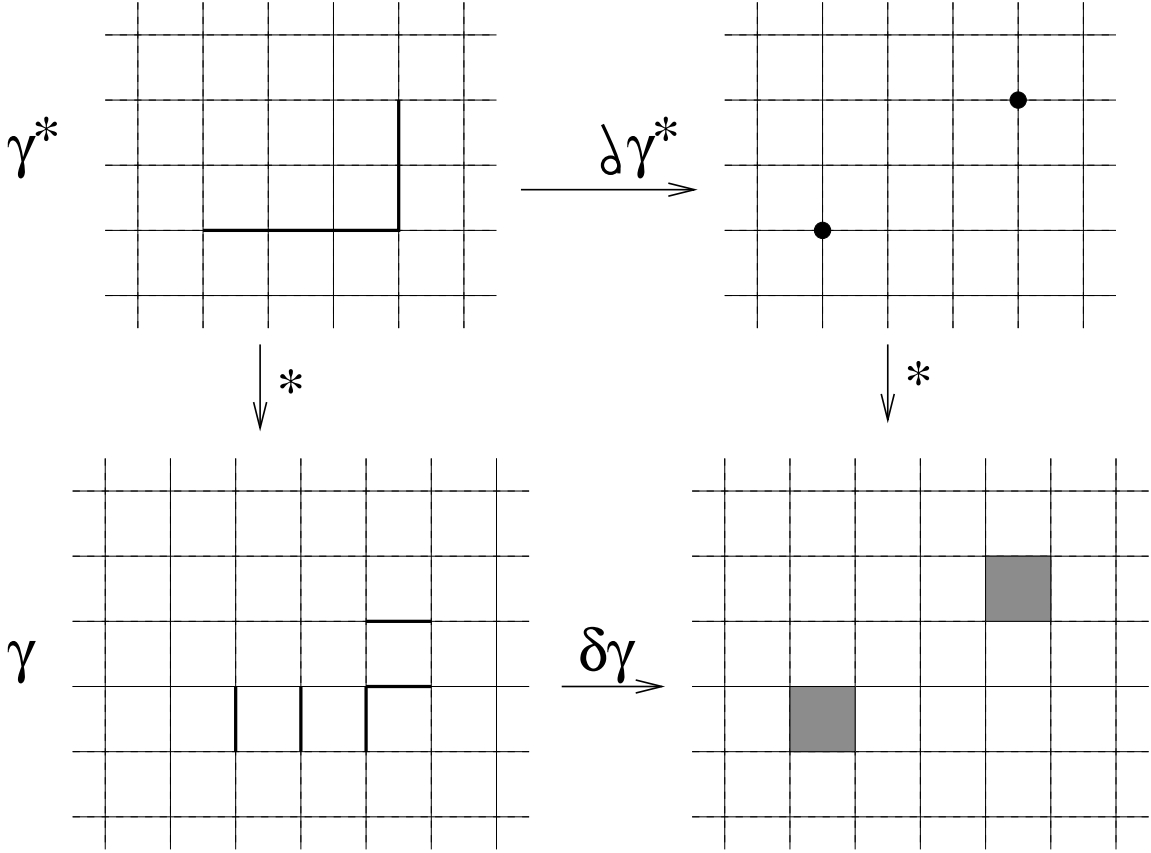}
\caption{The relation between disorder and order correlators in a 2-D spin 
system. The top diagram shows a curve which defines the 2-point
correlation function on the dual lattice, while the bottom diagram is
the collection of links which give the disorder correlator.
\label{disorderCorr2d}}
\end{figure}
As an example, consider the case of the Ising model in two
dimensions. A correlation function is labeled by a set of links which
start at site $i$ and end at site $j$ (see the top left diagram in
Figure \ref{disorderCorr2d}).  A disorder operator on the dual
lattice is obtained by interpreting the set of links joining site $i$
to site $j$ on the dual lattice. This corresponds to the links
displayed in the bottom left diagram in Figure
\ref{disorderCorr2d}. Inserting the characters along each link in that
diagram into the partition function computes the correlator of the two
disorder operators shown in the diagram on the bottom right. Notice
that they are dual to the two site variables in the top right.

\begin{figure}
\epsfxsize=6in
\epsfbox[0 0 561 137]{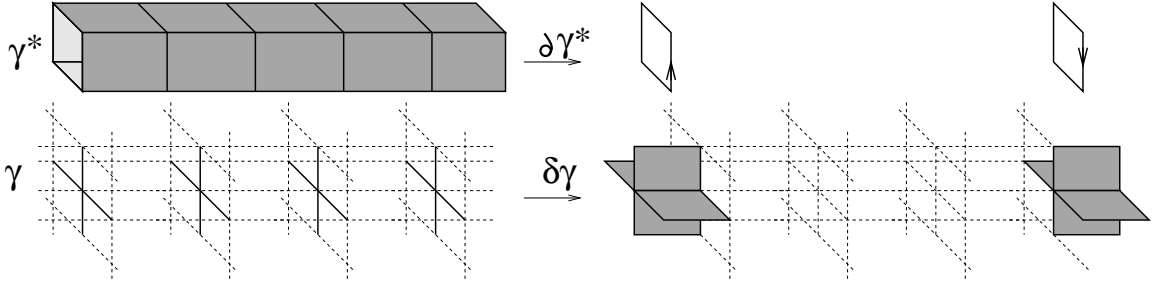}
\caption{The relation between disorder correlators in a 3-D spin system
and Wilson loops in a 3-D gauge theory. The top diagram is the surface
which defines the two Wilson loops at its ends, and the lower diagram
is its dual - a collection of links whose coboundary produces the
disorder variables in the spin model.
\label{disorderCorr3d}}
\end{figure}
As a second example consider a spin model in three dimensions. The
correlator is as usual given by a set of links joining site $i$ and site
$j$. A disorder correlator on the other hand is obtained by consider a
set of plaquettes forming a surface on the dual lattice. We will take
the example of a cylindrical like surface (see the top left diagram  in Figure
\ref{disorderCorr3d}). Interpreting the surface back on the original lattice 
leads to a set of links forming cross like shapes (see the bottom left
diagram in Figure \ref{disorderCorr3d}). Inserting the set of
characters for these links into the partition function computes the
correlator of the set of disorder variables shown in the bottom right
diagram of Figure \ref{disorderCorr3d}. Once again they are dual to
the correlator on the dual lattice, which in this case is the
correlator of two Wilson loops.

In the remainder of this section we will demonstrate how the duality
relations found in section \ref{DualModel} relates order and disorder
operators. We will show that there is a class of correlators which
vanish identically on topological grounds alone. Duality is in fact
even more powerful, and in certain dimensions it allows one to obtain
the correlation function of an arbitrary set of operators
explicitly. Several explicit examples will be explored at the end of
the section.

\subsection{General Formalism} \label{dualcorr}

In the previous section the correlators were defined by inserting
characters, in a particular fixed representation, on a collection of
$k$-cells, $\gamma$.  It is however unnecessary to fix the
representations on all of the cells of $\gamma$, one can allow the
representations to vary over the $k$-cells of $\gamma$.  In fact, it
is possible to compute such correlators explicitly in $d=k$
dimensions. That computation will be carried out shortly after the
general case has been worked out.

Here, we will consider a general correlator defined by an arbitrary
collection of $k$-cells denoted by $\gamma$ and a set of
representations $\{S_p \in \G^*\}$ for all $p\in\gamma$. These
correlators include, but are not limited to the order and disorder
correlators mentioned earlier. The correlator is defined much like
(\ref{correlator}),
\bea
W(\{S_p\}) &\equiv& \left\langle\prod_{p\in \gamma} \chi_{S_p} 
\left( \langle \delta \sigma, c_k^{(p)}\rangle\right)\right\rangle  
\nn\\
&=& \frac 1{Z}   \sum_{ \{n_a \in \H_a\} }
\sum_{\s \in C_{k-1}(\Omega, \G) } \prod_{p\in \gamma} \chi_{S_p} 
\left( \langle \delta \sigma, c_k^{(p)}\rangle\right)
\prod_{p=1}^{\N_k} B_p \left( \left \langle \delta \s+ n_a h^a,c_k^{(p)}
\right \rangle \right ) \label{DefCorr}
\eea
where $Z$ is of course the partition function (\ref{MODEL}).
The dual transformations on this correlator can be carried out in a
rather trivial manner. Firstly, let $\gamma_{s}$ denote the set of
$k$-cells with representation $s$, so that $\cup_s
\gamma_s = \gamma$. If $\gamma_{s}$ has no boundary the
correlation function of $\gamma_{s}$ reduces to the partition
function. This can be seen from (\ref{ordersimp}). Consequently, we
restrict the discussion to the case where $\gamma_{s}$ has a boundary
for every $s$. In that case, the product over characters can be
trivially absorbed in a redefinition of the Boltzmann weight,
\bea
B_p\left(\langle \delta \sigma +n_a h^a, c_k^{(p)} \rangle\right)
&\to& 
B_p\left(\langle \delta \sigma +n_a h^a, c_k^{(p)} \rangle \right) 
\chi_{S_p}\left( \langle \delta\sigma, c_{k}^{(p)} \rangle \right)\nn\\
&=&
\frac{1}{|\G|}\sum_{r_p\in \G^*} b_p(\langle r, c_k^{(p)}\rangle)
\chi_{\langle r, c_k^{(p)}\rangle} 
(\langle \delta\sigma +n_a h^a, c_{k}^{(p)} \rangle ) 
\chi_{\langle S, c_k^{(p)}\rangle}(\langle  \delta\sigma, c_{k}^{(p)} \rangle) 
\nn\\
&=& \frac{1}{|\G|}\sum_{r_p\in \G^*} b_p(\langle r-S, c_k^{(p)}\rangle)
\chi_{\langle r, c_k^{(p)}\rangle} (\langle \delta\sigma +n_a h^a, c_{k}^{(p)} \rangle ) \label{shiftBW}
\eea
for all $p\in \gamma$ and we have defined the $k$-chain $S \equiv\sum_s S_s =
\sum_{p\in\gamma} S_p\; c_k^{(p)}$. The last equality in (\ref{shiftBW}) was 
obtained by using the factorization properties of the characters,
(\ref{CharFactProp}), and then shifting $r$. Although the two
arguments of the characters in the second line appear to be different,
this difference vanishes for $p\in\gamma$ because one can always
arrange for the generators of the cohomology to have no overlap with
the cells of $\gamma$. To be precise $\langle S, h^a \rangle = 0$
since a non-vanishing inner product implies that $S$ contains an
element of the homology group, but this contradicts the assumption
that all $\gamma_{s}$ have a boundary.

With this shift in the Boltzmann weights the dual correlator is
trivial to obtain, simply use (\ref{Zdual}) with the appropriate
character coefficients.
\beq
W(\{S_p\}) = \frac{1}{Z} \sum_{\{\n_a\in \G^*\perp\H_a^*\}} \;
\sum_{r\in C_{d-k-1}(\Omega^*, \G^*)} \; 
\prod_{p=1}^{\N^*_{d-k}} b_p\left(\left\langle \delta r - S^*+ \n_a h^{*a},
c_{d-k}^{*(p)}\right\rangle \right) \label{DefCorrDual}
\eeq
The normalizing partition function, $Z$, should be evaluated in its
dual form (\ref{Zdual}) so that the factors of the group volume that
were left out to avoid clutter throughout, do not spoil the equality.
Also, $S^*$ is the dual to $S$, obtained by interpreting the set of
$k$-cells, $\gamma,$ on the original lattice as a set $d-k$ cells,
$\gamma^*$, on the dual lattice and defining $S^* =
\sum_{p\in\gamma^*} S_p\;c_{d-k}^{*(p)}$. It should be clear that if
$\gamma$ was chosen to define an order correlator, then
(\ref{DefCorrDual}) is a disorder correlator in the dual variables
thus establishing the connection between strong and weak coupling, and
order and disorder variables.

There is a class of correlators which this formalism does not
include. These correlators correspond to inserting characters along a
set of $(k-1)$-cells which form a closed sheet wrapping around a
non-trivial $(k-1)$-cycle of the lattice. For example, a gauge
theory on $S^1 \times {\cal M}$, where ${\cal M}$ is a contractable
space, has a Wilson loop which wraps once around the compact
direction. This set of links is not the boundary of a two-dimensional
sheet and hence cannot be written in the form
(\ref{DefCorr}). However, in the next section we will demonstrate 
that such correlators vanish identically.
\subsection{The Vanishing Correlator}
 
We now prove a powerful theorem, concerning the vanishing of certain
correlators in a statistical model. The physical interpretation of the
theorem will be explained, and some interesting applications will be
given shortly.

{\bf Theorem:} {\it Let $\gamma$ be a collection of $(k-1)$-cells with
zero boundary and $\{S_l\in\G^*\}$ a collection of representations
labeled by $l\in\gamma$. If $S\equiv \sum_{l\in\gamma} S_l\,
c_{k-1}^{(l)}$ is a representative element of $H_{k-1}(\Omega, \G^*)$ 
then the following correlator vanishes identically, $$
W(\gamma)\equiv\left\langle
\prod_{l\in \gamma} \chi_{S_l} \left( \sigma_l
\right)\right\rangle= 
\frac{1}{Z} \sum_{\{n_a\in\H_a\}}\sum_{\sigma \in C_{k-1}(\Omega, \G)}\;
\prod_{l\in\gamma}
\chi_{S_l} \left( \langle \sigma, c_{k-1}^{(l)} \rangle \right) 
\; \prod_{p=1}^{\N_k} \; B_p\left( \langle \delta \sigma + n_a h^a, c_{k}^{(p)}
\rangle \right) \label{VanCorr}
$$
where $Z$ is given by Eq.(\ref{MODEL})}. \\ \indent
{\bf Proof:} Introduce the character expansion of the Boltzmann weights, so
that every $k$-cell carries an irreducible representation, $\G^*$, of the 
group $\G$. Encode this information into a $k$-chain denoted by $r$ and
re-order the sum. The result is,
\bea
W(\{S_p\})\hspace{-2mm} &=&\hspace{-4mm}\sum_{r\in C_k(\Omega, \G^*)}\;\prod_{p=1}^{\N_k} b_p(
\langle r , c_k^{(p)} \rangle ) \nn\\
&&
\sum_{\{n_a\in\H_a\}}\;\sum_{\sigma \in C_{k-1}(\Omega, \G)}
\;\prod_{l\in\gamma} \chi_{S_l} \left( \langle \sigma, c_{k-1}^{(l)} \rangle 
\right) \; \prod_{p=1}^{\N_k} \; \chi_{\langle r, c_k^{(p)}\rangle}
\left( \langle \delta \sigma + n_a h^a , c_{k}^{(p)} \rangle \right)
\label{Wirrep}
\eea
where $b(r)$ are the character coefficients of the Boltzmann weights
given in (\ref{CharCoeff}). Using (\ref{bFlip}) and the factorization
properties of the characters, (\ref{CharFactProp}), one finds that the
sum over $\sigma$ and $\{n_a\}$ reduces,
\bea
\sum_{n_a, \sigma} \dots  
&=& \sum_{\sigma\in C_{k-1}(\Omega, \G)} \;
\prod_{l\in\gamma} \; \chi_{S_l} \left( \langle \sigma, c_{k-1}^{(l)}\rangle
\right )\; \prod_{l=1}^{\N_{k-1}} \; \chi_{\langle \partial r , c_{k-1}^{(l)}
\rangle} \left( \langle \sigma, c_{k-1}^{(l)} \rangle \right)
\sum_{\{n_a\in\H_a\}}\prod_{a=1}^{A^k} \chi_{\langle \partial r , h^a \rangle} 
\left( n_a \right)
\nn\\
&=& \sum_{\sigma\in C_{k-1}(\Omega, \G)} \;
\prod_{l=1}^{\N_{k-1}} \; \chi_{\langle \partial r + S, c_{k-1}^{(l)}
\rangle} \left( \langle \sigma, c_{k-1}^{(l)} \rangle \right)
\prod_{a=1}^{A^k}\sum_{n_a\in\H_a}\chi_{\langle \partial r , h^a \rangle} 
\left( n_a \right) \nn\\
&=& \prod_{l=1}^{\N_{k-1}} \; \delta_{\G^*} \left( \langle \partial r + S , 
c_{k-1}^{(l)} \rangle \right)
\prod_{a=1}^{A^k} \delta_{\H_a^*} \left(\langle \partial r , h^a \rangle 
\right)
\eea
where we have used the orthogonality of the characters,
(\ref{CharOrtho}), to obtain the last equality. If $S$ is a
representative element of $H_{k-1}(\Omega, \G^*)$ then the first delta
function is forced to vanish, since by definition there exists no
$k$-chain $r$ whose boundary equals an element of the homology group. Thus,
the correlator (\ref{VanCorr}) vanishes
$\Box$.

\begin{figure}
\hspace{30mm}\epsfxsize=4in
\epsfbox[0 0 418 166]{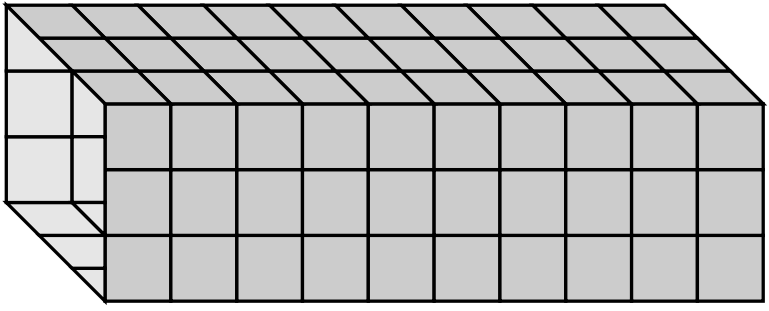}
\caption{This diagram depicts the set of plaquettes which are used in the
computation of the correlator between the two Wilson loops on its boundary.
\label{filledWLoop}}
\end{figure}
This rather general theorem has some interesting physical
consequences.  Consider for example a spin model ($k=1$) in arbitrary
dimensions and compute the correlator of an odd number of
characters. This vanishes identically from symmetry reasons alone,
however, from the point of view of the above theorem it vanishes since
an odd number of points is homologous to a single point, and hence is
a representative element of $H_0(\Omega,\G^*)$ and thus must vanish on
topological grounds. A more interesting application of the theorem is
to consider a gauge theory ($k=2$) on a manifold which has one compact
direction, and compute the Wilson loop around that compact direction
in representation $s$. This loop is clearly a representative of
$H_1(\Omega,\G^*)$ and thus the Wilson loop correlator must vanish
identically.  Although this result can be reasoned from symmetry
arguments, it seems somewhat more illuminating to see how it follows
from purely topological requirements. It is interesting to note that
if one were to compute the correlator of two Wilson loops wrapped once
around the compact direction with conjugate representations (i.e. one
with representation $s$ and the other with representation $-s$) it
does not vanish. The reason is because the $k$-chain
$S=s(\sum_{l\in\gamma_1} - \sum_{l\in
\gamma_2})$ is the boundary of a sheet formed between the two loops,
$S = s\;\partial
\sum_{p\in\Gamma} c_{2}^{(p)}$ where $\Gamma$ is set of plaquettes
connecting one loop to the other (see Figure \ref{filledWLoop}), and
is hence not a representative of $H_1(\Omega, \G^*)$. In fact, since
the correlator is reduced to a correlator for plaquette valued objects
one can use the results of section \ref{dualcorr} to complete the
duality transformation. We mention this since if the distance between
the two loops is taken to zero such a correlator corresponds to the
computation of a single Wilson loop correlator in the adjoint
representation which in a non-Abelian theory does not vanish. In
general there are only two fundamentally different types of
correlators, those that can be written in terms of the boundary of a
higher dimensional surface, which we have shown how to compute in the
previous section, and those that form representative elements of the
homology group, which has been demonstrated to vanish.

\subsection{Correlators in $d=k$ dimensions} \label{Corrdk}

Since we have established that a non-vanishing correlator is of the
general form (\ref{DefCorr}), and its dual formulation
(\ref{DefCorrDual}), in this section we deal only with those cases. It
is possible to obtain an explicit form for these correlators in
$d=k$ dimensions solely in terms of the topological properties of the
lattice. The reason is as follows: in performing the duality
transformations the constraints (\ref{rConstr}) must be solved,
however, in dimensions $d=k$ the most general solution to the first
set of constraints in (\ref{rConstr}) is $r=h\in H_d(\Omega, \G^*)$
because there are no $(d+1)$-chains on a $d$-dimensional
lattice. Consequently, the dual theory contains only topological
fluctuations. This sort of trivialization of the problem can also be
seen from the point of the view of the field-strength formulation
(\ref{FSform}). In dimensions $d=k$ the local Bianchi constraints are
absent rendering the dynamical fields locally free while constraining
them only globally. Of course the form of the dual correlator
(\ref{DefCorrDual}) implies this trivialization as well, since when
$d=k$ the dynamical field $\sigma$ is a $(-1)$-chain, which of course
has only the identity element, the only sums that remain are those
over $\{\n_a\}$. It is trivial to write down the correlators in
these dimensions, and they consists of only a finite number of sums
(or integrals if the dual group is continuous),
\beq
W(\{S_p\}) = 
\frac
{\sum_{\{\n_a\in \G^*\perp\H_a^*\}} \prod_{p=1}^{\N^*_{0}} 
b_p\left(\left\langle\n_a h^{*a} - S^*,c_{0}^{*(p)}\right\rangle \right)}
{\sum_{\{\n_a\in \G^*\perp\H_a^*\}} \prod_{p=1}^{\N^*_{0}} 
b_p\left(\left\langle\n_a h^{*a},c_{0}^{*(p)}\right\rangle \right)}
\label{dkCorrFn}
\eeq
here $\{h^{*a}\}$ are the set of generators of $H^0(\Omega^*,\G^*)$
which are dual to the generators of the homology group
$H_d(\Omega,\G^*)$. In the next two sections we will use this formula
to compute some non-trivial correlators in a spin model on an
arbitrary graph and a gauge theory on an arbitrary orientable
two-dimensional Riemann surface. These results have straightforward
generalizations to the higher dimensional cases.

\subsection{Spin System Case}

\begin{figure}
\epsfxsize=6in
\epsfbox[0 0 533 83]{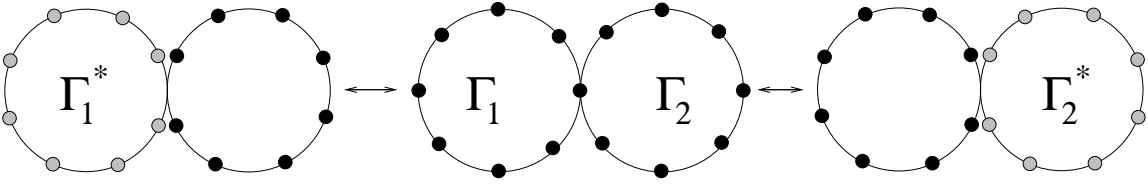}
\caption{A one dimensional graph and its dual. The middle diagram depicts the
original lattice, and $\Gamma_i$ are the set of links contained in the
corresponding loop. The end diagrams depict the dual lattice, and the
shaded sites are the duals to $\Gamma_i$, $\Gamma_i^*$,which generate
$H^0(\Omega^*,\G^*)$.
\label{graph1d}}
\end{figure}
In this section the computation of correlators in a spin model on a
$1$-dimensional graph using (\ref{dkCorrFn}) will be carried out.  Let
$\Omega$ be a one dimensional graph, and let $\{h^{*a}\}$
denote the generators of $H^0(\Omega^*, \G^*)$. To attach a physical
meaning to these generators consider the generators of $H_1(\Omega,
\G^*)$ and then re-interpret them on the dual lattice. It should be
obvious that $H_1(\Omega, \G^*)$ is generated by the set of links
which form closed loops on the graph (take only an independent set of
these generators). Figure \ref{graph1d} is an illustrative example
of how this works.  In general the generators of $H^0(\Omega^*,\G^*)$
are $h^a = \sum_{i\in\Gamma^*_a} C_0^{(i)}$ where $\Gamma^*_a$
($a=1,\dots,A$) is a collection points, on the dual lattice, which are
dual to a set of links that form a closed loop on the original
lattice. It is easy to convince oneself that the coboundary of $h^a$
vanishes, and since there are no lower dimensional chains this is an
element of the cohomology.  As long as the set of loops are
independent then these generators form a complete set.

The correlation function is defined via a set of links on the original
lattice, $\gamma$, and representations, $\{S_l\}$, for every link.
When $\gamma$ is interpreted on the dual lattice it corresponds to a
set of points, $\gamma^*$. Let $\gamma^*_a$ denote the set of points
in $\gamma^*$ that are contained in $\Gamma^*_a$ (i.e. the set of
points dual to the set of links contained in the $a$-th loop of the
original lattice). Also, let $\{S_i^*\}$ denote the collection of
representations on the dual sites, and define the $0$-chain ${\cal
S}_a^* \equiv \sum_{i\in\gamma^*_a} S^*_i \;c_0^{*(i)}$.  Applying the
dual formulation of the correlator (Eq.(\ref{dkCorrFn})), represented
by $\gamma$ gives,
\beq
W(\{S_p\}) = \prod_{a=1}^A \left\{
\frac{\sum_{\n_a\in \G^*\perp\H_a^*} \prod_{i\in\Gamma_a^*} 
b_i\left(n_a -\left\langle \S_a^*,c_{0}^{*(i)}\right\rangle \right) }
{\sum_{\n_a\in \G^*\perp\H_a^*} \prod_{i\in\Gamma_a^*}
b_i\left(\n_a \right)} \right\} \label{Spin1}
\eeq
We have made the assumption that the generators have no overlap, i.e.
$\Gamma_a \cap \Gamma_b = \emptyset$ if $a\ne b$. It is not difficult
to include the cases where there are overlaps, however, this only
serves to complicate the final sums over $\{n_a\}$. It is possible to
simplify the summations in the above expression only once the
Boltzmann weights are specified. A convenient example is furnished by
choosing $\G=U(1)$ and the Villan form for the Boltzmann weight,
so that,
\beq
b_i(\n) =\frac1{\sqrt{2\pi\beta}} \exp\left\{-\frac{1}{2\beta}\n^2\right\}
\label{Villan}
\eeq
The sums over $\n_a$ can now be performed in terms of well known
special functions, the Jacobi Theta function \cite{SpOl87}.  To
perform the sum it is also necessary to specify $\{\H_a\}$. Since $\G$
was taken to be $U(1)$ this forces $\H_a$ to be a cyclic group,
$\Z_{N_a}$. Inserting the above ansatz into Eq.(\ref{Spin1}) one finds,
\bea
\lefteqn{W(\{S_p\})} \nn\\
&&\hspace{-6mm}=\prod_{a=1}^A \left[
\exp\left\{- \frac1{2\beta} \sum_{i\in\gamma^*_a} S_i^2\right\}
\frac{\sum_{\n \in \Z} \exp\left\{ - (2\beta)^{-1} N^*_{0,a} N^2_a \n^2
+ \beta^{-1} N_a \n \sum_{i\in\gamma_a^*} S_i \right\}} {\sum_{\n \in
\Z} \exp\left\{ - (2\beta)^{-1} N^*_{0,a} N^2_a \n^2\right\}}
\right] \nn\\
&&\hspace{-6mm} =\prod_{a=1}^A\left[
\exp\left\{
- \frac1{2\beta}\left( \sum_{i\in\gamma_a^*} S_i^2 -
\frac{1}{\N_{0,a}^*}\left (\sum_{i\in\gamma_a^*} S_i\right)^2
\right)\right\}
\frac{\theta_3\left( \frac{2\beta}{N_a^2 \N_{0,a}^*}; -\frac{1}{N_a \N_{0,a}^*}
\sum_{i\in\gamma_a^*} S_i^* \right)}{\theta_3\left( \frac{2\beta}{N^2 
\N_{0,a}^*};0 \right)}
\right] 
\eea
Here $\N_{0,a}^*$ denotes the number of sites on the dual lattice
contained in $\Gamma_a^*$. This general formula is not extremely
illuminating, however, it reduces to a reasonable form for the two
point function. In that case $\gamma$ consists of a set of links which
form a single continuous path from point $i$ to $j$ on the lattice and
the representations on those links are all taken to be labeled by
$s$. Then the above expression reads,
\beq
W(\{s\}) = \prod_{a=1}^A\left[
\exp\left\{
- \frac{s^2}{2\beta} \; \frac{L(\gamma^*_a) ( \N_{0,a}^* -
L(\gamma^*_a))}{\N_{0,a}^*}
\right\}\frac{\theta_3\left( \frac{2\beta}{N_a^2 \N_{0,a}^*}; -
\frac{s\;L(\gamma_a^*)}{N_a \N_{0,a}^*} \right)} {\theta_3\left( \frac{2\beta}
{N_a^2 \N_{0,a}^*};0\right)} \right]
\eeq
where $L(\gamma_a^*)$ is the number of dual sites (links) contained in
$\gamma_a^*$ ($\gamma_a$) and counts the number of links that the
path $\gamma$ crossed in going through the $a$-th loop of the
graph. Notice that the overall exponential factor is explicitly
invariant under,
\beq
L(\gamma_a^*) \to \N_{0,a}^* - L(\gamma_a^*) \label{inout}
\eeq
This symmetry implies that the overall factor is not affected by which
path one took in going through a loop.  However, the theta-function
contributions are invariant only under an integer shift in the second
parameter. Write $s= s_1 + N s_2$ where $s_1\in \{ 0,\dots,N-1\}$,
$s_2 \in \Z$ and $N$ is the lowest common factor of $\{N_a\}$. It is
clear that under the transformation (\ref{inout}), the theta function
is invariant only if $s_1=0$. In general for a fixed $s_2$ there are
$N$ inequivalent sectors of the model, each sector transforms
differently under the transformation (\ref{inout}). The appearance of
these sectors can be traced back to the discussion in section
\ref{BIimp}. It was pointed out there that performing the sum over
topological sectors, $\{n_a\}$, in a subgroup of the gauge group
introduces fractional global charge into the system. In computing the
two-point function one is forced to introduce a set of links forming a
path with two end points. Naively, the choice of path has no effect on
the correlator, however, due to the existence of fractional charge in
the system there are lines of flux which pierce through the
path. Consequently, a different choice of path carries a different
amount of fractional charge, leading to a non-trivial transformation
of the correlator under the transformation (\ref{inout}). Only when
the representation of the path respects the $n$-ality of the global
charges can the system be symmetric under (\ref{inout}). This
requirement forces the representation $s$ to be an integer multiple of
the lowest common multiple of the global charges which it is exposed
to.

\subsection{Gauge Theory Case}
 
We now consider an Abelian gauge theory ($k=2$), with gauge group $\G$,
on an arbitrary orientable two-dimensional Riemann surface. Let
$\Omega$ be a discretetization of this manifold.  Since the generator
of the second homology group is the entire surface, this implies
that $H^0(\Omega,\G^*)$ has one generator which is given by
$h=\sum_{i\in\Omega^*} c_0^{*(i)}$. The correlator (\ref{DefCorr}) then
reduces to,
\beq
W(\{S_p\}) = \frac
{\sum_{\n \in \G^*\perp\H^*} \prod_{i=1}^{\N^*_{0}} b_i\left(\n-
\left\langle S^*,c_{0}^{*(i)}\right\rangle \right)} {\sum_{\n\in
\G^*\perp\H^*} \prod_{i=1}^{\N^*_{0}} b_i\left( \n \right)} \label{gaugeCorr}
\eeq
As in the last section consider $\G=U(1)$, $\H=\Z_N$ and the Boltzmann
weight in Villan form (\ref{Villan}).  Then Eq.(\ref{gaugeCorr}) reduces to,
\beq
W(\{ S_p\}) = \exp\left\{ - \frac{1}{2\beta} \left( \sum_{i\in\gamma^*}
(S_i^*)^2 - \frac{1}{\N_0^*} \left( \sum_{i\in\gamma^*} S_i^* \right)^2
\right)
\right\} \frac{ \theta_3\left( \frac{2\beta}{N^2 \N_0^*}; -\frac{1}{N \N_0^*}
\sum_{i\in\gamma^*} S_i^* \right)}{\theta_3\left( \frac{2\beta}{N^2 \N_0^*};0
\right)} \label{GeneralWilson}
\eeq
The simplest correlator in a gauge theory is a (filled) Wilson loop,
in which $\gamma$ is a collection of adjacent plaquettes and
$\{S_p=s\}$ (the two loops in Figure \ref{filledWLoop} is one such
example). Equation (\ref{GeneralWilson}) then becomes,
\beq
W(\{s\})
=
\exp\left\{-\frac{s^2}{2\beta}\;\frac{A(\gamma^*)(\N_0^*-A(\gamma^*))}{\N_0^*}
\right\} \frac{\theta_3\left( \frac{2\beta}{N^2 \N_0^*}; -\frac{s~A(\gamma^*)}
{N \N_0^*}
\right)} {\theta_3\left( \frac{2\beta}{N^2 \N_0^*}; 0\right)} \label{fWCF}
\eeq
Here $A(\gamma^*)$ denotes the number of dual sites (plaquettes)
contained in $\gamma^*$ ($\gamma$). Just as in the spin model case the
exponential is invariant under the following transformation,
\beq
A(\gamma^*) \to \N_0^* -A(\gamma^*) \label{inoutA}
\eeq
This symmetry  implies that the exponential part of the correlation
function does not distinguish between what is consider the inside or
outside area of the Wilson loop. Notice that in the limit in which the
area of the surface becomes infinite while the area of the Wilson loop
remains finite ($A(\gamma^*)/\N_0^* \to 0$) the exponential reduces to
the familiar area law. However, if the ratio of the number of
plaquettes in the Wilson loop and the total number of plaquettes on
the manifold is kept fixed ($A(\gamma^*)/\N_0^* ={\rm const.}$) as the
number of plaquettes is taken to infinity, so that one reproduces the
continuum limit, then there are finite size corrections given both by
the theta-function and the overall exponential.

As in the spin model scenario, the theta-function does not respect the
symmetry (\ref{inoutA}) in general.  The generic case has $s = s_1 + N
s_2$ where $s_1\in \{ 0,\dots,N-1\}$ and $s_2\in\Z$, if $s_1\ne 0$ the
system distinguishes between inside and out of the loop, just as in
the one-dimensional spin model. The reasoning follows through much
like it did there. The fractional charge induces a flux which is
incompatible with the representation (unless $s=N\Z$) and as such a
different choice of surface carries a different amount of flux. The
$N$ different sectors are once again seen to give rise to $N$
different transformations of the correlator under (\ref{inoutA}). This
bears a strong resemblance to the appearance of theta sectors in
non-Abelian gauge theories. It indicates that there is a strong
connection between theta sectors in a non-Abelian theory, and the
topological sectors in an appropriate Abelian model on a topologically
non-trivial manifold.

Before closing we would like to illustrate how one can include direct
product groups in a rather trivial manner. As an example consider the
case in which the gauge group $\G=\oplus_{a=1}^m U(1)$ and $\H =
\oplus_{a=1}^m \Z_{N_m}$. Also choose the heat kernel action, $$
b_i((\n^1,\dots, \n^m) ) = \prod_{a=1}^m\frac{1}{\sqrt{2\pi\beta_a}}
\exp \left \{ -  \frac 1 {2\beta_a} (n^a)^2 \right\}
$$ here $\{\beta_a\}$ are independent coupling constants.  Since the
gauge group is a product group it is necessary to specify multiple
representations of $U(1)$ on every plaquette in $\gamma$, label these
representations by $\{S_p^a\}$.  Since the Boltzmann weights have no
cross terms the problem factorizes and the correlation function is
straightforward to write down, $$ W(\{ S^a_p\}) =
\prod_{a=1}^{m} \left[
\exp\left\{ - \frac{1}{2\beta_a} \left( \sum_{i\in\gamma^*}
(S^{*a}_i)^2 - \frac{1}{\N_0^*} \left( \sum_{i\in\gamma^*} S^{*a}_i \right)^2
\right) \right\} 
\frac{ \theta_3\left( \frac{2\beta_a}{N^2_a \N_0^*}; -\frac{1}{N_a \N_0^*}
\sum_{i\in\gamma} S^{*a}_i \right)}{\theta_3\left( \frac{2\beta_a}{N^2_a \N_0^*}
;0\right)}\right] $$ Once again consider the simplifying case of a
filled Wilson loop in which $\{S_p^a = s\}$ for all $p\in\gamma$. The
above expression then reduces to a product of factors like that
appearing in (\ref{fWCF}),
\beq
W(\{s\}) =
\exp\left\{-\frac{s^2}{2{\tilde\beta}
\N_0^*}\;A(\gamma)(\N_0^*-A(\gamma))\right\}\prod_{a=1}^{m} \left( 
\frac{\theta_3\left( \frac{2\beta_a}{N^2_a \N_0^*}; -\frac{s~A(\gamma)}
{N_a \N_0^*} \right)} {\theta_3\left( \frac{2\beta_a}{N^2_a \N_0^*}; 0\right)}
\right)
\eeq
where ${\tilde \beta} = \sum_{a=1}^m \beta_a^{-1}$.  The invariance of
this expression under $A(\gamma) \to \N_0^* - A(\gamma)$ depends upon
whether $s$ is a representation which respects the $n$-ality of the
fractional charge. In this case only if $s\in N\Z$ where $N$ is the
least common multiple of $\{N_a\}$ will the expression be invariant
under that symmetry.

\section{Conclusions} \label{conclusion}

We have introduced some new statistical models in which a sum over
various topological sectors of the system were included. The dual
model was constructed explicitly, and several useful examples were
presented, making connection with the usual notations and
conventions. Using this duality transformation the model was written
in terms of the natural field-strength variables which were found to
be constrained by local and global Bianchi identities. This
formulation of model was derived in a gauge-invariant manner and the
sum over the topological sectors were seen to introduce fractional
charges along the non-trivial $k$-cycles of the lattice. The fact that
it is possible to introduce a sum over different groups for the
different $k$-cycles was used to form self-dual models in which there
were several distinct global charges in the system. These are new
self-dual models and the analysis of \cite{DaHa97} could be used to
identify the self-dual points in these models as possible new critical
points. Finally, the duality transformations were carried out on
correlation functions in various dimensions. We proved that a certain
subset of correlators vanished identically, in addition it was shown
how correlators in dimensions $d=k$ are completely determined by the
duality transformation in terms of the topological modes of the
theory.

\section{Acknowledgments}

The author would like to thank T. Fugleberg, L. D. Paniak,
O. Tirkkonen, G. W. Semenoff and A. Zhitnitsky for useful discussions
on various aspects of this paper. The hospitality of the Niels Bohr
Institute, where part of this work was completed, was also much
appreciated.

%
% --------------------------------------------------------
%

\end{document}